\documentclass[a4paper,10pt]{article}
\usepackage[T1]{fontenc}
\usepackage[utf8]{inputenc}
\usepackage[english]{babel}
 
\usepackage{amsmath}
\usepackage{amssymb}
\usepackage{amsthm}
\usepackage{enumitem}
\usepackage{appendix}
\usepackage{geometry}
\usepackage[colorlinks=true,linkcolor=blue,urlcolor=blue]{hyperref}
\usepackage{eurosym}
\usepackage{hyperref}
\hypersetup{
    colorlinks,
    citecolor=blue,
    filecolor=blue,
    linkcolor=blue,
    urlcolor=blue
}
\usepackage{todonotes}
\usepackage{mathrsfs}

 \usepackage{rotating}
 \usepackage[round]{natbib}

\def\be{\begin{eqnarray}}
\def\ee{\end{eqnarray}}

\def\b*{\begin{eqnarray*}}
\def\e*{\end{eqnarray*}}

\newcommand{\ba}{\begin{array}}
\newcommand{\ea}{\end{array}}
\newcommand{\ben}{\begin{equation*}} 
\newcommand{\een}{\end{equation*}}
\newcommand{\bea}{\begin{eqnarray}}
 \newcommand{\eea}{\end{eqnarray}}
\newcommand{\bean}{\begin{eqnarray*}} 
\newcommand{\eean}{\end{eqnarray*}}
\newcommand{\bel}{\begin{align}} 
\newcommand{\eel}{\end{align}}
\newcommand{\beln}{\begin{align*}} 
\newcommand{\eeln}{\end{align*}}
\newcommand{\bit}{\begin{itemize}}
\newcommand{\eit}{\end{itemize}}

\makeatletter \@addtoreset{equation}{section}

\@addtoreset{Definition}{section}

\@addtoreset{Theorem}{section}

\@addtoreset{Proposition}{section}

\@addtoreset{Property}{section}

\@addtoreset{Assumption}{section}

\@addtoreset{Corollary}{section}

\@addtoreset{Lemma}{section}

\@addtoreset{Remark}{section}

\@addtoreset{Example}{section}

\@addtoreset{Condition}{section}

\def \E{\mathbb{E}}

\def \H{\mathbb{H}}
\def \L{\mathbb{L}}
\def \M{\mathbb{M}}
\def \N{\mathbb{N}}
\def \P{\mathbb{P}}
\def \Q{\mathbb{Q}}
\def \R{\mathbb{R}}

\def \Z{\mathbb{Z}}

\def \G{\mathbb{G}}

\def\={\;=\;}
\def\.{\;.}

\def\1{{\bf 1}}

 \def\normeL2#1{\left\|{#1}\right\|_{L^2}}
 
\newcommand{\alias}[2]{
\providecommand{#1}{}
\renewcommand{#1}{#2}
}

\alias{\P}{\mathbb{P}}
\alias{\N}{\mathcal{N}}
\alias{\L}{\mathcal{L}}
\alias{\Z}{\mathbb{Z}}
\alias{\Q}{\mathbb{Q}}
\alias{\R}{\mathbb{R}}
\alias{\C}{\mathcal{C}}
\alias{\T}{\mathbb{T}}
\alias{\E}{\mathbb{E}}
\alias{\H}{\mathcal{H}}
\alias{\1}{\mathbf{1}}
\alias{\B}{\mathcal{B}}
\alias{\M}{\mathcal{M}}
\alias{\G}{\mathcal{G}}
\alias{\Y}{Y_{\bullet}}

\newcommand{\nc}{\newcommand}
\nc{\cA}{{\mathcal A}} \nc{\cB}{{\mathcal B}} \nc{\cC}{{\mathcal
C}} \nc{\cD}{{\mathcal D}} \nc{\bbD}{\mathbb{D}}
\nc{\cG}{{\mathcal G}} \nc{\cF}{{\mathcal F}} \nc{\cS}{{\mathcal
S}} \nc{\cU}{{\mathcal U}} \nc{\cH}{{\mathcal H}}
\nc{\cK}{{\mathcal K}} \nc{\cM}{{\mathcal M}} \nc{\cO}{{\mathcal
O}} \nc{\cP}{{\mathcal P}} \nc{\bbE}{\mathbb{E}}
\nc{\bbEP}{\mathbb{E}_{\mathbb{P}}}\nc{\bbL}{\mathbb{L}}
\nc{\bbP}{\mathbb{P}} \nc{\bbQ}{\mathbb{Q}} \nc{\del}{\partial}
\nc{\Om}{\Omega} \nc{\om}{\omega} \nc{\bbR}{\mathbb{R}}
\nc{\bbC}{\mathbb{C}} \nc{\bfr}{\begin{flushright}}
\nc{\efr}{\end{flushright}} \nc{\dXt}{\delta q_{t}}
\nc{\dXs}{\delta q_{s}} \nc{\bs}{\blacksquare} \nc{\dX}{\delta q}
\nc{\dY}{\Delta Y}
\nc{\dnkx}{\left(X(T^{n}_{k})-X(T^{n}_{k-1})\right)}
\nc{\esssup}{\mathrm{ess}\mbox{ }\mathrm{sup}}
\nc{\essinf}{\mathrm{ess}\mbox{ } \mathrm{inf}}
\nc{\dhats}{\widehat{\delta_s}}

\nc{\chf}{\mbox{$\mathbf1$}}
\nc{\ind}{\mathds{1}}

\begingroup\expandafter\expandafter\expandafter\endgroup
\expandafter\ifx\csname pdfsuppresswarningpagegroup\endcsname\relax
\else
\fi
\graphicspath{{./fig/}}

\nc{\mum}{ \mu_{\rm m} }
\nc{\muv}{ \mu_{\rm v} }
\nc{\mumv}{ \mu_{\rm mv} }
\nc{\Hm}{ H_{\rm m} }
\nc{\Hv}{ H_{\rm v} }

\usepackage{import}

\usepackage[onehalfspacing]{setspace}
\usepackage{dsfont}
\newcommand{\uproman}[1]{(\uppercase\expandafter{\romannumeral#1})}

\usepackage{bbm}
\newcommand{\abstand}{\hspace{0pt}}

\usepackage{graphicx}
\usepackage{float}

\usepackage{booktabs}
\usepackage{ragged2e}
\usepackage{array}
\newcolumntype{C}[1]{>{\centering\let\newline\\\arraybackslash\hspace{0pt}}m{#1}}

\usepackage{footnote}

\usepackage{caption}
\usepackage{subcaption}
\graphicspath{{Figures_19-May-2022/Arismendi/}{Figures_19-May-2022/Samuelson/}{Figures_19-May-2022/FanelliSchmeck/}{Figures_19-May-2022/Comparison/}{Figures_19-May-2022/Options/}{Figures_19-May-2022/}}  

\usepackage{tikz}
\usetikzlibrary{automata,arrows,positioning,calc,shapes, backgrounds,fit, chains, scopes,shapes.geometric}

\newtheorem{prop}{Proposition}
\newtheorem{theorem}{Theorem}
\newtheorem{remark}{Remark}
\newtheorem{lemma}{Lemma}

\begin{document}
\title{The Market Price of Risk for Delivery Periods:\\
	Pricing Swaps and Options in Electricity Markets}

\author{Annika Kemper\thanks{Center for Mathematical Economics (IMW) at Bielefeld University. Financial support from the Deutsche Forschungsgemeinschaft (DFG, German Research Foundation) – SFB 1283/2 2021 – 317210226 is gratefully acknowledged.}  
	\quad Maren D. Schmeck\thanks{Center for Mathematical Economics (IMW) at Bielefeld University.
		Financial support from the Deutsche Forschungsgemeinschaft (DFG, German Research Foundation) – SFB 1283/2 2021 – 317210226 is gratefully acknowledged.} 
	\quad Anna Kh. Balci\thanks{Faculty of Mathematics at 
		Bielefeld University. 
		Financial support from the Deutsche Forschungsgemeinschaft (DFG, German Research Foundation) – SFB 1283/2 2021 – 317210226 is gratefully acknowledged.}}
\maketitle
\maketitle
\begin{abstract}
In electricity markets, futures contracts typically function as a swap since they deliver the underlying over a period of time. In this paper, we introduce a market price  for the delivery periods of electricity swaps, thereby opening an arbitrage-free pricing framework for derivatives based on these contracts.
Furthermore, we  use  a weighted \textit{geometric averaging} of an artificial geometric futures price over the corresponding delivery period. Without any need for approximations, this averaging results in geometric swap price dynamics. Our framework allows for including typical features as the Samuelson effect, seasonalities, and  stochastic volatility. 
In particular, we investigate the pricing procedures for electricity swaps and options in line with \cite{Arismendi2016}, \cite{schneider2018samuelson}, and \cite{fanelli2019seasonality}. 
A numerical study highlights the differences between these models depending on the delivery period.


\end{abstract}




\section{Introduction}
Futures contracts are  the most important  derivatives in electricity and commodity markets. 
Due to the non-storability of electricity, the underlying is typically delivered over a period, and the contract is therefore referred to as a swap. 
In electricity markets, the delivery period has an influence on price dynamics, and \cite{fanelli2019seasonality} have provided empirical evidence indicating that implied volatilities of electricity options are seasonal with respect to the delivery period. 
In other words, the distributional features -- or the pricing measure -- depend on the delivery period of the contract.
In this paper, we introduce an arbitrage-free pricing framework that takes dependencies on the delivery into account. The core of our approach is the so-called \textit{market price of delivery risk}, which reflects expectations about variations in volatility weighted over the delivery period and arises through a \textit{geometric average} approach similar to that used by \cite{KemnaVorst1990}.

\begin{figure}
	\centering
	\includegraphics[width=0.65\columnwidth]{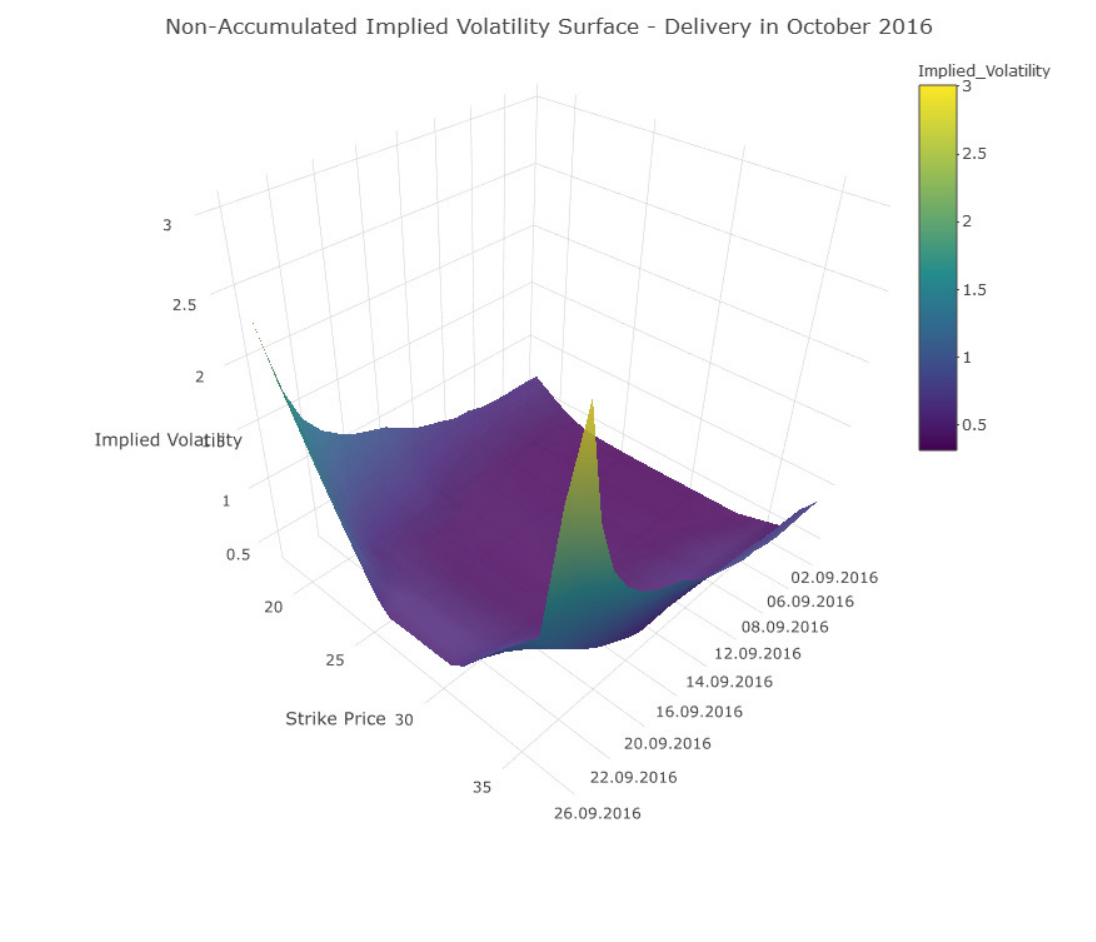}
	\caption[Caption for LOF]{The implied non-accumulated volatility surface with respect to strikes from 18 to 38 over the last trading month in September 2016 for a European call option on the Phelix DE/AU Baseload Month futures at the European Energy Exchange (EEX) delivering in October 2016.}
	\label{fi:implvolasurface}
\end{figure}

In fact, the delivery period is one of the features that distinguishes electricity markets from other commodity markets such as oil, gas, or corn.  An easy way to acknowledge its existence is to use futures price dynamics with a delivery time that represents the midpoint of the delivery period. This approach has been followed, for example, by \cite{schmeck2016}, and is advantageous as it captures the typically observed behavior that the futures prices do not converge against the electricity spot price if time approaches the beginning of the delivery. A possible way to model the delivery period explicitly is to average the spot price or an artificial futures price over the entire delivery time. Typically, \textit{arithmetic averaging} is used, which is the standard approach in electricity price modeling   and  works especially well for arithmetic price dynamics (see, e.g., \cite{Benth2008}, and \cite{Benth2019}).
However, if the underlying electricity futures are of the geometric type, the resulting dynamics are neither geometric  nor Markovian.
In that case, the dynamics are approximated in line with \cite{Bjerksund} (see also \cite{Benth2008}). \\
A typical feature of electricity markets is the seasonal behavior of prices. The effect is enforced through the rise of renewable energy, which is highly dependent on weather conditions. At  present, there is a growing worldwide trend  to acknowledge the need for sustainable energy production, which also raises  the expectations of a further increasing impact of seasonal effect. Among others, \cite{Arismendi2016}, \cite{BorSchmeck}, and \cite{fanelli2019seasonality} have addressed and modeled seasonality in either commodity or energy markets. Typically, a deterministic seasonal price level is added to the price dynamics, but the dynamics can also exhibit seasonal behavior.    
\cite{fanelli2019seasonality} distinguish between \textit{seasonalities in the trading day} and \textit{seasonalities in the delivery period}. 
\cite{Arismendi2016} suggest the use of a seasonal stochastic volatility model for commodity futures. As in the Heston model, stochastic volatility follows a square-root process, but with a seasonal mean-reversion level.
Indeed, a volatility smile can also be observed in electricity option markets (see Figure \ref{fi:implvolasurface}) such that a stochastic volatility model seems appropriate.\\
Finally, a well-known feature in electricity and commodity markets is the Samuelson effect (see \cite{Samuelson}), which implies that futures close to delivery are much more volatile than are those whose expiration date lies far off.
This effect can be observed in the implied volatility of electricity options, especially  far out and in the money  (see also Figure \ref{fi:implvolasurface} and  \cite{kiesel} ).
The effect is typically included in any electricity futures price dynamics.  \cite{schneider2018samuelson} include such a term-structure effect within the framework of stochastic volatility modeling.  \cite{schmeck2016} investigates analytically the impact of the Samuelson effect on option pricing.

In this paper, we suggest modeling the delivery period explicitly through a \textit{geometric averaging} approach for electricity futures prices of the geometric type, in line with \cite{KemnaVorst1990} and \cite{Bjerksund}. This approach  leads directly to Markovian and geometric swap price dynamics. Indeed, the
geometric averaging of futures prices coincides with the arithmetic procedure applied to logarithmic futures prices.
In line with the literature,  we base the averaging procedure on an artificial futures contract that is a martingale under a  pricing measure $\mathbb{Q}$. In our framework, the resulting swap price dynamics are not a martingale under $\mathbb{Q}$   due to a  drift term in the dynamics that is characterized by the variance of the weighted delivery and is
 used to define  the market price of delivery risk and an equivalent martingale measure $\widetilde{\mathbb{Q}}$ for the swap price.  $\widetilde{\mathbb{Q}}$ can thus be used as a pricing measure for derivatives on the swap. We characterize the market price of delivery risk for the Samuelson effect, and for seasonalities in the trading day and in the delivery period following \cite{schneider2018samuelson}, \cite{Arismendi2016}, and \cite{fanelli2019seasonality}, respectively. \\
For option pricing, we consider a general stochastic volatility model that is inter alia feasible for mean-reverting square-root volatility processes in line with the models used by \cite{Arismendi2016} and \cite{schneider2018samuelson}.
The volatility structure is rich enough to include the categories of seasonalities and the Samuelson effect.
Both models share the feature that their commodity futures prices are based on an affine stochastic volatility structure.
Indeed, the averaging procedure of the futures price model as well as the change of measure preserve the affine model structure of the artificial futures price dynamics.\\
In this paper, we focus on the pricing of a single swap contract. As mentioned above, the pricing measure depends on this particular contract, and it thus cannot be used for pricing derivatives on another swap contract with a different delivery period. Nevertheless,  several swap contracts are usually also tradable, such that arbitrage possibilities must be excluded.  Furthermore, overlapping delivery periods are tradable as a quarter and the corresponding three months. We address how to tackle these issues in Section \ref{sec:Arbitrage_Cons}. 

The paper is organized as follows: 
Section~\ref{sec:averaging} presents the geometric averaging approach and introduces the market price of delivery risk based on a general stochastic volatility model.
In order to illustrate the averaging procedure, we discuss the method based on the models created by \cite{Arismendi2016},  \cite{schneider2018samuelson}, and \cite{fanelli2019seasonality} in Section \ref{sec:Trafo_Models}.
In Section \ref{sec:Arbitrage_Cons}, we address how to exclude arbitrage opportunities that might appear when there are several, possibly overlapping swap contracts traded on the market. Option pricing is discussed in  Section~\ref{sec:options}.
In addition, all adjusted commodity market models are investigated numerically. 
Finally, Section~\ref{sec:summary} presents our conclusions.

\section{Averaging of Futures Contracts} \label{sec:averaging}
We consider a swap contract delivering  a  flow of 1 Mwh electricity during the delivery period $(\tau_1,\tau_2]$. 
At a trading day $t\leq \tau_1$,  the swap price is denoted by $F(t, \tau_1,\tau_2)$ and settled such that the contract is entered at no cost. 
It can be interpreted as an average price of instantaneous delivery. Motivated by this interpretation, we consider an artificial futures contract with price $f(t,\tau)$ that stands for instantaneous delivery at time $\tau\in(\tau_1,\tau_2]$. Note that such a contract does not exist on the market, but turns out to be useful for modeling purposes when considering delivery periods (see for example \cite{Benth2019}).

Consider a filtered probability space $(\Omega, \mathcal{F}, (\mathcal{F}_t)_{t\in[0,\tau]}, \mathbb{Q})$,
where the filtration satisfies the usual conditions.
At time $t\leq\tau$, the price of the futures contract follows a geometric diffusion process evolving as
\begin{align} 
&df(t,\tau) = \sigma(t,\tau) f(t,\tau)dW^{f}(t)\;, \label{eq:futurespricedynamicsonetime}\\
&d\sigma^2(t,\tau)=a(t,\tau, \sigma)dt+c(t,\tau, \sigma)dW^{\sigma}(t)\;,\label{eq:futurespricedynamicsonetime2}
\end{align}  
with initial conditions $f(0,\tau) = f_0 >0$ and $\sigma^2(0,\tau)=\sigma^2_0>0$, and
where  $W^{f}$ and $W^{\sigma}$ are correlated standard Brownian motions under $\mathbb{Q}$. 
Thus, $W^{\sigma}=~\rho W^{f}+~\sqrt{1-\rho^2}W$ for a Brownian motion $W$ independent of $W^f$ and $\rho\in(-1,1)$. We assume that both, the futures price volatility $\sigma(t, \tau)$
and the futures price $f(t,\tau)$, are  $\mathcal{F}_t$-adapted for $t\in[0,\tau]$, and that they satisfy
suitable integrability and measurability conditions  to ensure that  \eqref{eq:futurespricedynamicsonetime} is a $\mathbb{Q}$-martingale, and the solution given by
\begin{align} \label{eq:SolutionFutures}
f(t,\tau)=f(0,\tau)e^{\int_{0}^{t}\sigma(s,\tau)dW^f(s)-\frac12 \int_{0}^{t}\sigma^2(s,\tau)ds}
\end{align}
exists  (see Appendix \ref{app:tech_requ} for details).
As $\sigma(t,\tau)$ depends on both time $t$ and delivery time $\tau$, we allow for volatility structures as the Samuelson effect, seasonalities in the trading day, or seasonalities in the delivery time. In this framework, we would like to mention the models of \cite{Arismendi2016}, \cite{schneider2018samuelson}, as well as of \cite{fanelli2019seasonality}, which are addressed in the next section.
 
Following the Heath-Jarrow-Morton approach to price futures and swaps in electricity markets, the swap price is usually defined as the \textit{arithmetric average} of futures prices (see, e.g., \cite{Benth2008},  \cite{Bjerksund}, and \cite{Benth2019}):
\begin{equation} \label{eq:arithmeticnoarbitrage}
F^a(t, \tau_1, \tau_{2}) = \int_{\tau_1}^{\tau_2}w(u, \tau_1, \tau_{2})f(t,u)du\;,
\end{equation} for a general weight function
\begin{align}
w(u,\tau_1, \tau_2):=\frac{\hat{w}(u)}{\int_{\tau_1}^{\tau_2}\hat{w}(v)dv}\;, ~~~~ \text{for } u\in(\tau_1,\tau_2]\;.
\end{align}
The most popular example is given by  $\hat{w}(u) =1$, such that $w(u,\tau_1, \tau_2)= \frac{1}{\tau_2 - \tau_1}$. This corresponds to a one-time settlement. A continuous settlement over the time interval $(\tau_1,\tau_2]$ is covered by  $	\hat{w}(u) =e^{-ru}$, where $r$ is the constant interest rate (see, e.g., \cite{Benth2008}).  The arithmetric average of the futures price as in  \eqref{eq:arithmeticnoarbitrage} leads to tractable dynamics for the swap as long as one  assumes an arithmetric structure of the futures prices as well. 
This is based on the fact that arithmetic averaging is tailor-made for absolute growth rate models.
Nevertheless,  if one defines the futures price as a geometric process as in \eqref{eq:futurespricedynamicsonetime}, one can show that the dynamics of the swap defined through \eqref{eq:arithmeticnoarbitrage} is given by
\begin{align*}
dF^a(t,\tau_1, \tau_2) 
=\sigma(t,\tau_2)F^a(t,\tau_1,\tau_2)dW^f(t) -\int_{\tau_1}^{\tau_2}\frac{\partial \sigma}{\partial u}(t,u)\frac{w(\tau, \tau_1,\tau_2)}{w(\tau, \tau_1, u)} F^a(t, \tau_1,u) du ~dW^f(t)\;,
\end{align*} for any $\tau \in (\tau_1,\tau_2]$ (see \cite{Benth2008}; Chapter 6.3.1). Thus, the dynamics of the swap price is neither a geometric process nor Markovian, which makes it unhandy for further analysis.  \cite{Bjerksund} suggest an approximation  given by
\begin{align} \label{eq:arithAverage}
dF^a(t,\tau_1,\tau_2)=F^a(t,\tau_1,\tau_2) \Sigma(t,\tau_1,\tau_2)dW^f(t)\;,
\end{align}
where $F^a(0,\tau_1,\tau_2)=f_0$ 
and a weighted average volatility 
\begin{align} \label{eq:Sigma}
\Sigma(t,\tau_1, \tau_{2}) := \int_{\tau_1}^{\tau_2}w(u, \tau_1, \tau_{2})\sigma(t,u)du\;.
\end{align}
Instead of averaging absolut price trends as in \eqref{eq:arithmeticnoarbitrage}, we here suggest to
focus on the averaging procedure of  relative price trends, i.e. growth rates or logarithmic prices.  
This leads to a \textit{geometric averaging} procedure in continuous time. 
In fact, the connection between exponential models and geometric averaging seems natural:
the geometric averaging of a geometric price process corresponds to an arithmetic average of logarithmic prices.
Note that this approach is in line with  \cite{KemnaVorst1990} for pricing average asset value options on equities and also with \cite{Bjerksund}. The difference of \cite{Bjerksund} and our approach is, that \cite{Bjerksund} approximate the geometric average to receive a martingale dynamics, while we will make a change of measure. Note that the choice of  pricing measures in electricity markets allows for more freedom as in other markets, as electricity itself is not storable, and thus no-arbitrage considerations for the spot itself are not applicable (see \cite{BenthSchmeck2014b}).

We  define the swap price as 
\begin{align} \label{eq:GeomAverage}
F(t, \tau_1, \tau_{2}) := \exp\left(\int_{\tau_1}^{\tau_2}w(u, \tau_1, \tau_{2})\log(f(t,u))du\right)\;.
\end{align}
Assume that the volatility satisfies further integrability conditions  (see Appendix \ref{app:tech_requ}).
It turns out, that the resulting swap price dynamics is a geometric process with stochastic swap price volatility $\Sigma(t,\tau_1, \tau_{2})$:
\begin{lemma}
The dynamics of the swap price defined in \eqref{eq:GeomAverage} under $\mathbb{Q}$ are given by
\begin{equation*}\label{eq:FunderQ}
\begin{split}
\frac{d F(t,\tau_1,\tau_2)}{F(t,\tau_1,\tau_2)} =- 
\frac{1}{2}\left(\int_{\tau_1}^{\tau_2}w(u, \tau_1, \tau_{2})\sigma^2(t,u)du -\Sigma^2(t,\tau_1, \tau_{2}) \right) dt +\Sigma(t,\tau_1, \tau_{2})~dW^{f}(t)\;.
\end{split}
\end{equation*}
\end{lemma}
\begin{proof}
Plugging \eqref{eq:SolutionFutures} into \eqref{eq:GeomAverage} and using the stochastic Fubini Theorem (see \cite{Protter2005}; Theorem~65) leads to 
\begin{align}
F(t,\tau_1,\tau_2)
=F(0, \tau_1, \tau_{2})
e^{-\frac{1}{2}\int_{0}^{t}\int_{\tau_1}^{\tau_2}w(u, \tau_1, \tau_{2})\sigma^2(s,u)du~ ds+~\int_{0}^{t}\Sigma(s,\tau_1,\tau_2)~dW^{f}(s)}\;. \label{eq:av_price}
\end{align}
Then, the result follows using Itô's formula.
\end{proof}
Although the futures price process $f(\cdot,\tau)$ is a martingale under the pricing measure $\mathbb{Q}$, the swap price process $F(\cdot,\tau_1,\tau_2)$ is not a $\mathbb{Q}$-martingale anymore: the swap price dynamics under $\mathbb{Q}$ has a negative drift term, given by the difference between the weighted average of the futures price variance and the swap price's variance $\Sigma^2$.
We thus define a \textit{market price of delivery risk} at time $t\in[0,\tau_1]$ associated to the delivery period $(\tau_1, \tau_2]$ as 
\begin{align}\label{eq:marketpricerisk}
b_1(t,\tau_1, \tau_2):=-\frac{1}{2}\frac{\int_{\tau_1}^{\tau_2}w(u, \tau_1, \tau_{2})\sigma^2(t,u)du-\Sigma^2(t,\tau_1,\tau_2)}{\Sigma(t,\tau_1,\tau_2)}\;,
\end{align}
where $b_1(t,\tau_1, \tau_2)$ is measurable and $\mathcal{F}_t$-adapted as $\sigma(t,u)$ and $\Sigma(t,\tau_1,\tau_2)$ are. 
It can be interpreted as the trade-off 
between the variance of the swap $\Sigma^2$
on the one hand and the weighted average variance of a stream of futures on the other hand. Since we have two independent Brownian motions, $W^f$ and $W$, we have a two-dimensional market price of risk $b(t,\tau_1,\tau_2)=(b_1(t,\tau_1,\tau_2),b_2)^\intercal$, where we choose $b_2=0$.
The market price $b_1(\cdot,\tau_1,\tau_2)$ will enter also the dynamics of the volatility, which is driven by the Brownian motion $W^\sigma= \rho W^f+ \sqrt{1-\rho^2}W$.
 \begin{remark} \label{re:1}
	For a random variable $U$ with density $w(u,\tau_1,\tau_2)$, we can write
	\begin{align*}
	\Sigma(t,\tau_1,\tau_2)&= \mathbb{E}_U[\sigma(t,U)]\;, \quad \text{and } \quad
	b_1(t,\tau_1, \tau_2)=-\frac12 \frac{\mathbb{V}_U[\sigma(t,U)]}{\mathbb{E}_U[\sigma(t,U)]}\;,
	\end{align*}
	where $\mathbb{E}_U$ and $\mathbb{V}_U$ denote the expectation and variance only with respect to the random variable~$U$.
	Note that $\sigma(t,U)$ identifies the futures price volatility for a random time of delivery.
	Hence, the market price of delivery risk is the variance per unit of expectation of $\sigma(t,U)$.
	This is very similar to the well-known coefficient of variation $\frac{\sqrt{\mathbb{V}_U[\sigma(t,U)]}}{\mathbb{E}_U[\sigma(t,U)]}$.
\end{remark} 
We define a new pricing measure $\widetilde{\mathbb{Q}}$, such that $F(\cdot, \tau_1, \tau_{2})$ is a martingale.
Define the Radon-Nikodym density through
\begin{align*}
Z(t, \tau_1, \tau_2):=\exp\left\{- \int_{0}^{t} b_1(s,\tau_1,\tau_2)dW^f(s) -\frac12 \int_{0}^{t}b_1^2(s,\tau_1,\tau_2)ds\right\}\;.
\end{align*}
 Assume that 
\begin{align}\label{eq:mart}
\mathbb{E}_\mathbb{Q}\left[ Z(\tau_1,\tau_1,\tau_2) \right]=1\;,
\end{align}
which means $Z(\cdot, \tau_1, \tau_2)$ is indeed a martingale for the entire trading time. We will show later that Novikov's condition (see, e.g., \cite{KaratzasShreve1991}) is fulfilled for suitable models, such that \eqref{eq:mart} holds true.
We then define the new measure  $\widetilde{\mathbb{Q}}$ through the Radon-Nikodym density
\begin{align*}
\frac{d\widetilde{\mathbb{Q}}}{d\mathbb{Q}}:=Z(\tau_1, \tau_1, \tau_2)\;,
\end{align*}
which clearly depends on the delivery period $(\tau_1,\tau_2]$.
Girsanov's theorem  states that
\begin{align}\label{eq:Girsanov}
\widetilde{W}^f(t)=& W^f(t)+\int_0^t b_1(s,\tau_1, \tau_2)ds\;, \\
\widetilde{W}(t)=& W(t)\;,
\end{align}
are standard Brownian motions under $\widetilde{\mathbb{Q}}$ (see, e.g., \cite{Shreve2004}).
The Brownian motion of the stochastic volatility is also affected due to the correlation structure:
\begin{align}
 \widetilde{W}^\sigma(t) =W^\sigma(t)+\int_{0}^{t}\rho b_1(s,\tau_1, \tau_2)ds\;. \label{eq:measurechange_vola}
\end{align}
A straight forward valuation  leads to the following result:
\begin{prop}
The swap price process $F(\cdot,\tau_1,\tau_2)$ defined in \eqref{eq:GeomAverage} is  a martingale  under $\widetilde{\mathbb{Q}}$.
The swap price and volatility dynamics are given by
\begin{align}
\frac{d F(t,\tau_1,\tau_2)}{F(t,\tau_1,\tau_2)} =& \Sigma(t,\tau_1,\tau_2)~d\widetilde{W}^{f}(t)\;,\label{eq:FunderQtilde} \\
d\sigma^2(t,\tau)=&\left(a(t,\tau,\sigma)-\rho b_1(t,\tau_1,\tau_2)c(t,\tau,\sigma)\right)dt+c(t,\tau,\sigma)d\widetilde{W}^{\sigma}(t)\;,
\end{align}
where $\Sigma(t,\tau_1,\tau_2)$ is defined in \eqref{eq:Sigma}.
\end{prop}
Note that the stochastic volatility process $\sigma^2(t,\tau)$ also depends on the delivery interval, which we drop for notational convenience.
 As the swap price process $F(\cdot,\tau_1,\tau_2)$ is a martingale under the equivalent measure $\widetilde{\mathbb{Q}}$, we can use it to price options on the swap. Nevertheless,  $\widetilde{\mathbb{Q}}$ depends on the particular delivery period of the swap
and cannot be used to price options on swaps on other delivery periods. We address this issue in Section \ref{sec:averaging_multi}.

We would like to compare the approximated swap price $F^a(t,\tau_1,\tau_2)$ under $\mathbb{Q}$ following \cite{Bjerksund}  with the swap price $F(t,\tau_1,\tau_2)$ under $\widetilde{\mathbb{Q}}$ as defined in \eqref{eq:GeomAverage} assuming that both have a stochastic volatility based on \eqref{eq:futurespricedynamicsonetime2}. The swap price dynamics have the same form, the difference is in the drift term of the stochastic volatility. If the volatility is deterministic as in the setting of \cite{Bjerksund}, the distribution of $F^a(t,\tau_1,\tau_2)$ under $\mathbb{Q}$ and the distribution of $F(t,\tau_1,\tau_2)$ under $\widetilde{\mathbb{Q}}$ are the same. For the swap prices both under the same measure we have the following result.
\begin{lemma} For the swap prices $F^a(t,\tau_1,\tau_2)$  and  $F(t,\tau_1,\tau_2)$, both under $\mathbb{Q}$, it holds that
	\begin{align*}
	F(t,\tau_1,\tau_2)-F^a(t,\tau_1,\tau_2) 
	=&F^a(t,\tau_1,\tau_2)\left[e^{\frac12 \int_{0}^{t} \mathbb{V}_U\left[\sigma(s,U)\right]ds}-1\right]\geq 0\;.
	\end{align*}
\end{lemma}
 \begin{proof}
 	From \eqref{eq:arithAverage}, we know that
 	\begin{align*}
 	F^a(t,\tau_1,\tau_2)=f_0 e^{-\frac12\int_{0}^{t}\Sigma^2(s,\tau_1,\tau_2)ds+\int_{0}^{t}\Sigma(s,\tau_1,\tau_2)dW^f(s)}\;.
 	\end{align*} 
 	Using  equation \eqref{eq:av_price} and the notation from Remark \ref{re:1},  we find
 	\begin{align}\label{eq:neu}
 	F^a(t,\tau_1,\tau_2)=F(t,\tau_1,\tau_2)e^{-\frac12 \int_{0}^{t} \mathbb{V}_U\left[\sigma(s,U)\right]ds}
 	\end{align} 
 	and the result follows. The expression in squared brackets is strictly positive as it is the case for the variance.
 \end{proof}
Note that in \eqref{eq:neu}, $\mathbb{V}_U\left[\sigma(s,U)\right]$ can be interpreted as discount rate.

\section{Electricity Swap Price Models} \label{sec:Trafo_Models}
In this section, we transform three  commodity market models from the recent literature into electricity swap models using the geometric averaging procedure presented in Section \ref{sec:averaging}. 
All models induce separate effects each surrounded by a stochastic volatility framework to capture the volatility smile.
In Section \ref{sec:Arismendi}, we examine the 	influence of seasonality in the mean-reversion level of the (stochastic) volatility following  \cite{Arismendi2016}.
Section \ref{sec:TS} considers the impact of the Samuelson effect being empirically confirmed by \cite{kiesel} for electricity swaps.
The futures price model which will be transformed is in line with \cite{schneider2018samuelson} who apply their term-structure model to the agricultural futures market.
In Section \ref{sec:FS}, we investigate delivery-dependent seasonalities based on the empirical evidence by \cite{fanelli2019seasonality}. 
In particular, we suggest a futures price model inducing seasonalities in the delivery time and transform the model accordingly.
For all three models, we investigate the corresponding swap and market prices numerically.
Note, that the model by \cite{Arismendi2016} with constant mean-reversion level serves as a comparative benchmark, since the delivery time does not affect the volatility.  In Section \ref{sec:options}, we then address option pricing for these three models.

\subsection{Seasonal Dependence on the Trading Day} \label{sec:Arismendi} 
\begin{figure}[tb]
	\centering
	\begin{subfigure}[b]{0.45\textwidth}
		\centering
		\caption{Swap Price Evolution}
		\includegraphics[height=5.3cm]{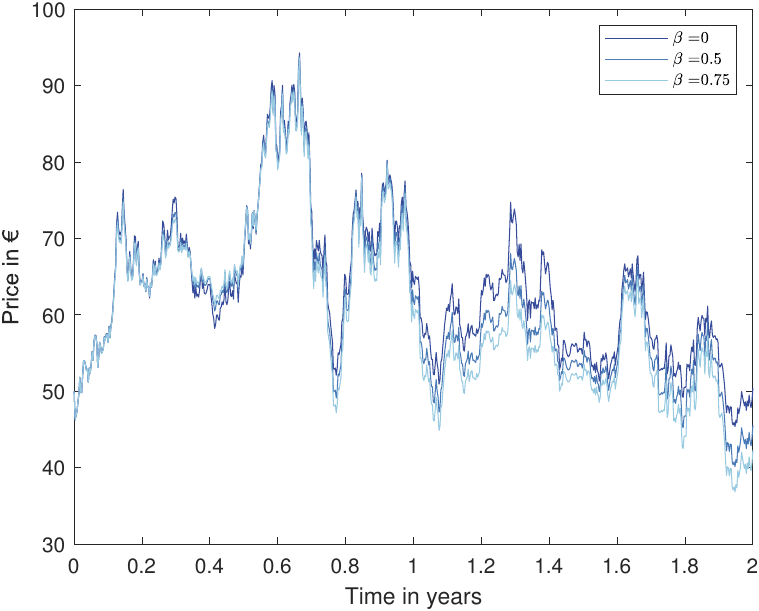}
	\end{subfigure}
	\hfill
	\begin{subfigure}[b]{0.45\textwidth}
		\centering
		\caption{Stochastic Volatility}
		\includegraphics[height=5.3cm]{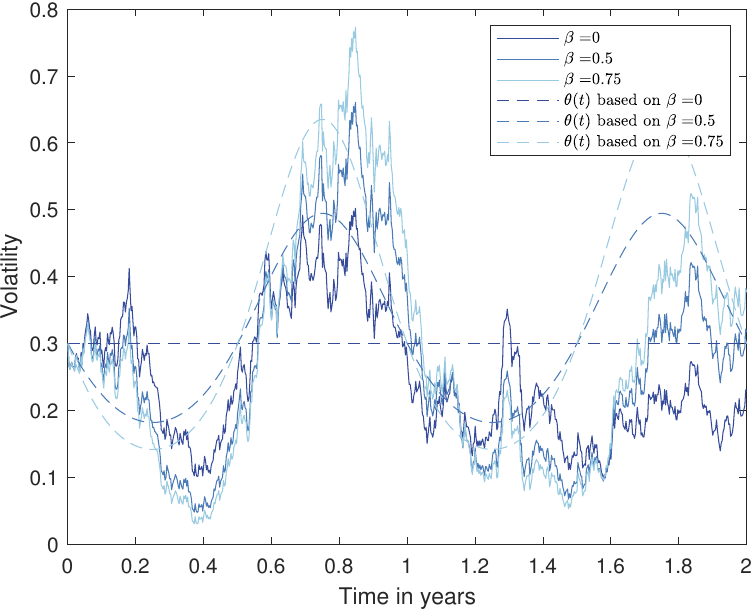}
	\end{subfigure}
	\caption{Seasonal dependence on the trading day. Subfigure (a) illustrates swap prices based on the stochastic volatilities for different amplitudes $\beta$ in the mean-reversion level of the stochastic volatility. Subfigure (b) presents the stochastic volatility for different amplitudes $\beta$ in its mean-reversion level. For the choice of parameters, see Tables \ref{tab:1} and \ref{tab:1_Arismendi}.}
	\label{fi:Vola_Swap_Arismendi_beta}
\end{figure}

\cite{Arismendi2016} consider a generalized Heston model, where the mean-reversion rate of the stochastic volatility is seasonal in trading time. That is, they suggest a futures price dynamics of the form
\begin{align}
df(t,\tau)=& \sqrt{\nu(t)}f(t,\tau)dW^f(t)\;, \label{eq:ArismendiF} \\
d\nu(t) =& \kappa\left(\theta(t)-\nu(t)\right)dt + \sigma\sqrt{\nu(t)}dW^{\sigma}(t)\;, \label{eq:vola_underQ}
\end{align}
where $W^{\sigma}$ and $W^f$ are defined as before under $\mathbb{Q}$.  The stochastic volatility $\nu(t)$ is given by  a  Cox-Ingersoll-Ross process with a time-dependent level. 
The Feller condition $2\kappa \theta^{\min} > \sigma^2$ needs to be satisfied with $\theta^{\min}:= \min_{t\in[0,\tau]}\theta(t)$
in order to receive a strictly positive solution.
If the mean-reversion level $\theta(t)$  is  in particular of exponential sinusoidal form, that is $\theta(t)=\alpha e^{\beta \sin(2\pi(t+\gamma))}$, for $\alpha,\beta>0, \gamma\in[0,1)$, then $\theta^{\min}=\alpha e^{-\beta}$.
In particular, $\alpha$ represents the long-term variance level of the stochastic volatility, strengthened or  mitigated by the seasonal exponential function driven by the amplitude $\beta$ and the shift $\gamma$ of the sine-function, where $\gamma=0$ identifies January 1 and where $\beta=0$ serves as a benchmark without seasonal effects.
In the framework of Section~\ref{sec:averaging},  the futures price volatility is given by $\sigma(t,\tau)=~\sqrt{\nu(t)}$.
The corresponding swap price dynamics under the $\mathbb{Q}$ evolve as
\begin{align}
	dF(t,\tau_1,\tau_2)=&\sqrt{\nu(t)}F(t,\tau_1,\tau_2)dW^f(t)\;, \label{eq:Trafo_Arismendi_F}\\
	d\nu(t)=& \kappa(\theta(t)-\nu(t))dt+\sigma\sqrt{\nu(t)}dW^\sigma(t)\;.\label{eq:Trafo_Arismendi_nu}
\end{align}
Typical trajectories of the volatility and swap prices are illustrated in Figure  \ref{fi:Vola_Swap_Arismendi_beta}.
As the futures price volatility does not depend on the delivery time $\tau$, the resulting volatility of the swap is given by the futures price volatility 
\begin{align}\label{eq:Vola_Arismendi}
\Sigma(t,\tau_1,\tau_2)= \sqrt{\nu(t)}\;,
\end{align} 
for all choices of weight functions $w(\cdot,\tau_1,\tau_2)$. 
Then, the market price of the delivery period is also zero, that is
\begin{align}
	b_1(t,\tau_1,\tau_2)=0\;,
\end{align}
for all $t\in[0,\tau_1]$ and we  arrive directly at swap price dynamics of martingale form. Since the model is not linked to the delivery time, the pricing measures for the futures and swap contract coincide, as  the dynamics do.  
In Figure \ref{fi:Vola_Swap_Arismendi_beta}, we illustrate the model for different amplitudes $\beta$ in the mean-reversion level of the volatility process.. The higher $\beta$, the more pronounced the mean-reversion level, and the higher the stochastic volatility oscillates (see Figure \ref{fi:Vola_Swap_Arismendi_beta} (b)). Hence, the higher $\beta$, the more the swap's trajectory fluctuates (see Figure \ref{fi:Vola_Swap_Arismendi_beta} a)).

\subsection{Samuelson Effect} \label{sec:TS}
\cite{schneider2018samuelson} include the so-called Samuelson effect within the framework of a futures price model under  stochastic volatility. The Samuelson effect describes the empirical observation that the variations of futures increase the closer the expiration date is reached (see also \cite{Samuelson}). Typically this is captured with an exponential alteration in the volatility of the form $ e^{-\lambda(\tau-t)}$, for $\lambda>0$. For $t\to \tau$, the term converges to $1$ and the full volatility enters the dynamics. If the time to maturity increases,  that is for $\tau-t\to \infty$,
the volatility decreases. While
\cite{schneider2018samuelson}  base their model on a multi-dimensional setting, 
we here focus on the one-dimensional case following
\begin{align}
df(t,\tau)=& e^{-\lambda(\tau-t)}\sqrt{\nu(t)}f(t,\tau)dW^f(t)\;, \label{eq:F_Q_TS}\\
d\nu(t)=& \kappa(\theta-\nu(t))dt+\sigma\sqrt{\nu(t)}dW^\sigma(t)\;. \label{eq:nu_Q_TS}
\end{align}
This approach includes a  term-structure in the volatiliy process of the form $\sigma(t,\tau)= e^{-\lambda(\tau-t)}\sqrt{\nu(t)}$.
Applying the geometric averaging method as in \eqref{eq:GeomAverage},
the volatility of the swap is 
\begin{align}
\Sigma(t,\tau_1,\tau_2)=d_1(\tau_2-\tau_1)e^{-\lambda(\tau_1-t)}\sqrt{\nu(t)}\;, \label{eq:Sigma_Samuelson}
\end{align}
 and the new swap martingale measure $\widetilde{\mathbb{Q}}$ is defined via the market price of risk 
\begin{align}\label{eq:b1samuelson}
 b_1(t,\tau_1,\tau_2)=-d_2(\tau_2 -\tau_1)e^{-\lambda(\tau_1-t)}\sqrt{\nu(t) }\;,
 \end{align}
where for $\hat{w}(u)=1$,
\begin{align}
 d_1(x)=\frac{1-e^{-\lambda x}}{\lambda x}\;, \quad\text{and}\quad  
 d_2(x)=\frac12\left(\frac12 (1+e^{-\lambda x})-d_1(x)\right)\;, \label{eq:d1_d2}
\end{align}
and for $\hat{w}(u)=e^{-ru}$ with $r\neq -\lambda$ and $r\neq -2\lambda$,
\begin{align}
d_1(x)=\frac{r}{r+\lambda} \frac{1-e^{-(r+\lambda)x}}{1-e^{-rx}}\;, \quad \text{and} \quad 
d_2(x)=\frac12\left(\frac{r+\lambda}{r+2\lambda}\frac{1-e^{-(r+2\lambda)x}}{1-e^{-(r+\lambda)x}}-d_1(x)\right)\;. \label{eq:d1_d2_contw}
\end{align}
Note that whenever $r\to 0$, then $\hat{w}(u)\to 1$ and observe that $d_1$ remains positive for $x>0$ and $d_2$ stays positive in the constant settlement case for $x>0$ and $\frac{2-\lambda x}{2+\lambda x}e^{\lambda x}<1$.
The volatility and the market price of risk factorize into three parts: a constant, $d_1(\tau_2-\tau_1)$ or $-d_2(\tau_2-\tau_1)$, respectively, depending only on the length of the delivery period, the Samuelson effect counting the time to maturity at $\tau_1$, and the stochastic volatility $\sqrt{\nu(t)}$.
The Samuelson effect enters both swap price dynamics and market price of delivery risk through the term  $ e^{-\lambda(\tau_1-t)}$.  $\Sigma$ (and $b_1$) become small (large) if we are far away from maturity, and increases (decreases) exponentially if we approach the maturity of the swap. 
The swap price dynamics under  $\widetilde{\mathbb{Q}}$  are  given by
 \begin{align}
dF(t,\tau_1,\tau_2)=&\Sigma(t,\tau_1,\tau_2)F(t,\tau_1,\tau_2)d\widetilde{W}^f(t)\;, \label{eq:F_TS}\\
d\nu(t) =& \left(\kappa\theta-\left[\kappa-\rho\sigma d_2(\tau_2-\tau_1)e^{-\lambda(\tau_1-t)}\right]\nu(t)\right)dt+\sigma\sqrt{\nu(t) }d\widetilde{W}^{\sigma}(t)\;, \label{eq:nu_TS}
\end{align}
where $d\widetilde{W}^\sigma(t) = dW^\sigma(t) + \rho b_1(t,\tau_1,\tau_2)dt$ using \eqref{eq:measurechange_vola}.
We observe that the drift of the dynamics of $\nu(t)$ is now altered by the market price of risk, which again depends on the delivery period. The speed of mean reversion is now given by $\kappa-\rho\sigma d_2(\tau_2-\tau_1)e^{-\lambda(\tau_1-t)}$.
In particular, if $\rho d_2(\tau_2-\tau_1)>0$, we need to assume that $\kappa>\sigma\rho d_2(\tau_2-\tau_1)$ such that the speed stays positive.
 The mean reversion level is given by $\frac{\kappa\theta}{\kappa-\rho\sigma d_2(\tau_2-\tau_1)e^{-\lambda(\tau_1-t)}}$.
For a negative correlation  between swap price and volatility dynamics and a positive \mbox{$d_2(\tau_2-\tau_1)$,} the speed of mean reversion increases under the new measure and the level of mean reversion decreases and vice versa for a positive correlation with positive $d_2(\tau_2-\tau_1)$. 
If \mbox{$\kappa^2> \sigma^2 d_2(\tau_2-\tau_1)^2$,} Novikov's condition is satisfied such that the measure change is well defined and the process $F(\cdot, \tau_1,\tau_2)$ is indeed a true martingale under $\widetilde{\mathbb{Q}}$ (see Appendix \ref{app:Novikov_TS}). 
Using the notation of Remark~\ref{re:1}, we can write
\begin{align*}
\Sigma(t,\tau_1,\tau_2)&= \mathbb{E}_U[e^{-\lambda(U-\tau_1)}]e^{-\lambda(\tau_1-t)}\sqrt{\nu(t)}\;,\\
 b_1(t,\tau_1,\tau_2)
 &=-\frac12 \frac{ \mathbb{V}_U[e^{-\lambda(U-\tau_1)}]}{\mathbb{E}_U[e^{-\lambda(U-\tau_1)}]}e^{-\lambda(\tau_1-t)}\sqrt{\nu(t)}\;,
\end{align*}
 for a random variable $U\sim \mathcal{U}[\tau_1, \tau_2]$.
\begin{figure}[tb]
\centering
\begin{subfigure}[b]{0.3\textwidth}
\centering
\caption{Swap Price Evolution}
\includegraphics[height=3.8cm]{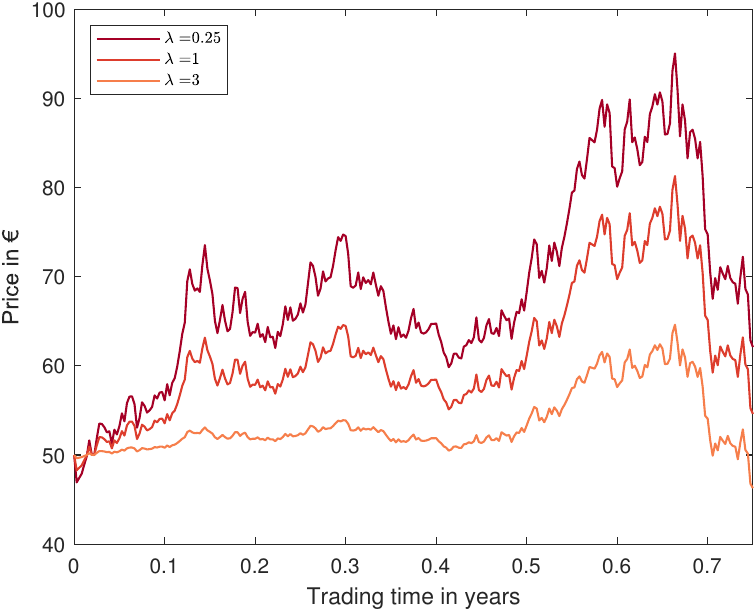}
\end{subfigure}
\hfill
\begin{subfigure}[b]{0.3\textwidth}
\centering
\caption{Swap Price Volatility}
\includegraphics[height=3.8cm]{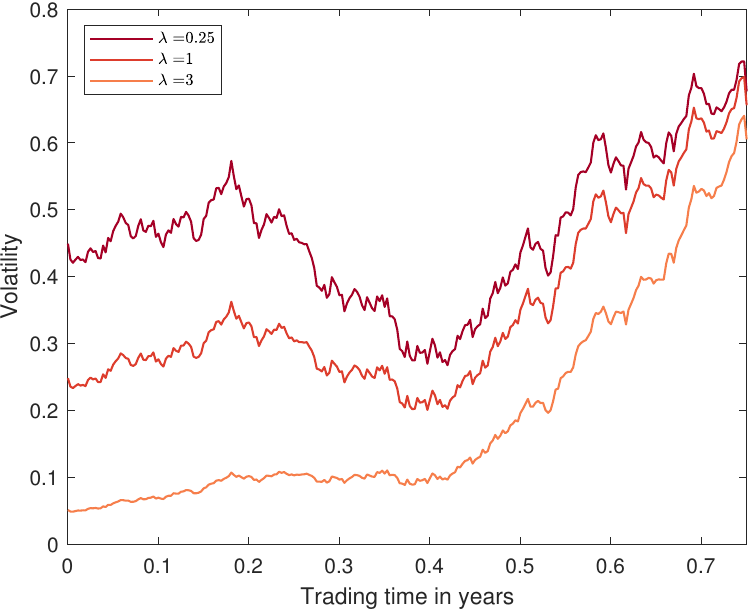}
\end{subfigure}
\hfill
\begin{subfigure}[b]{0.3\textwidth}
	\centering
	\caption{Market Price Delivery Risk}
	\includegraphics[height=3.8cm]{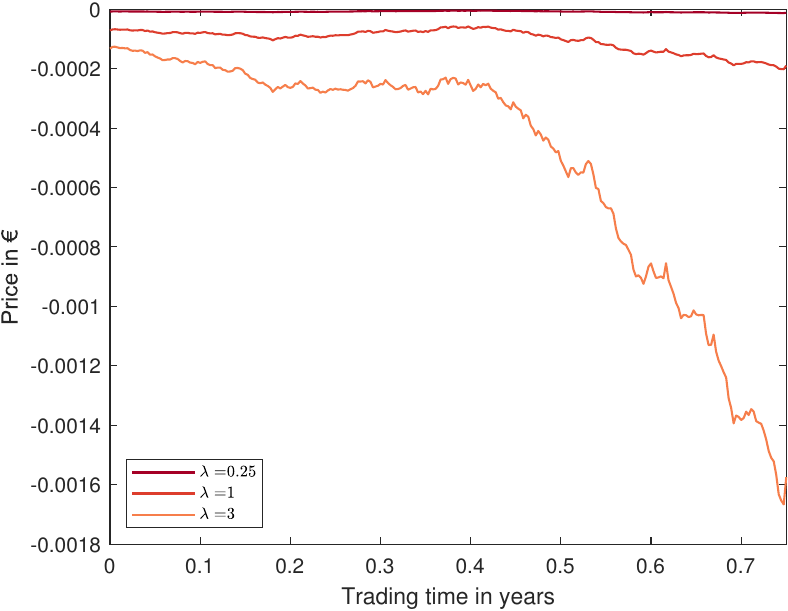}
\end{subfigure}
\caption{Samuelson effect under constant settlement. Subfigure (a) shows the swap prices under $\widetilde{\mathbb{Q}}$. Subfigure (b) presents the swap price volatility. Subfigure (c) presents the market prices of delivery risk. The parameter set is based on Tables \ref{tab:1} and \ref{tab:1_Samuelson} for different values $\lambda$.}
\label{fi:Vola_Swaps_TS_Different_lambdas}
\end{figure}
The impact of the Samuelson effect on the market price of delivery risk as well as the swap price and volatility dynamics is illustrated in Figure \ref{fi:Vola_Swaps_TS_Different_lambdas}.
The parameters are chosen as in Tables \ref{tab:1} and \ref{tab:1_Samuelson} (see Section \ref{sec:SimComp}). The exponential behavior of the market price becomes more pronounced the higher the Samuelson parameter $\lambda$ since $d_2(\frac{1}{12})$ increases (see Figure \ref{fi:Vola_Swaps_TS_Different_lambdas} (c) and Table \ref{tab:1_Samuelson}). At terminal time, it is equal to $-d_2(\frac{1}{12})\sqrt{\nu(\tau_1)}$, which depends by definition on $\lambda$ (see Equation \eqref{eq:b1samuelson} and Table \ref{tab:3}).
Moreover, we clearly observe the Samuelson effect within the swap price evolution.
The higher the Samuelson parameter, the smaller the expectation of the Samuelson effect, $d_1(\frac{1}{12})$, and the higher the variance of the Samuelson effect (see Table \ref{tab:3}). Hence, with increasing Samuelson parameter, the swap's volatility becomes lower. 
However, the closer we reach the expiration date, the higher the swap price volatility (see Figures \ref{fi:Vola_Swaps_TS_Different_lambdas} (a) and \ref{fi:Vola_Swaps_TS_Different_lambdas} (b)). 
\begin{table}[tb]
	\centering
	\begin{tabular}{C{30mm} C{30mm} C{30mm} C{30mm}}
		\midrule
		&\multicolumn{1}{c}{$d_1(\frac{1}{12})=\mathbb{E}_U\left[e^{-\lambda (U-\tau_1)}\right]$} & \multicolumn{1}{c}{$\mathbb{V}_U\left[e^{-\lambda (U-\tau_1)}\right]$}  & \multicolumn{1}{c}{$d_2(\frac{1}{12})$} \\
		 \cmidrule{1-4} 
		\midrule
		$\lambda = 0.25$ & $0.9897$  &$1.7712e-05$ & $1.7897e-05$ \\[1ex]
		$\lambda = 1$ & $0.9595$  &$0.0005$ & $0.0003$ \\[1ex]
		$\lambda = 3$ & $0.8848$  &$0.0041$ & $0.0023$ \\[1ex]
		\midrule
	\end{tabular}%
\caption{Expectation, variance and market price of delivery risk for different Samuelson parameters based on the parameter values in Tables \ref{tab:1} and \ref{tab:1_Samuelson}.}
\label{tab:3}%
\end{table}%

\subsection{Delivery-Dependent Seasonality} \label{sec:FS}
 \cite{fanelli2019seasonality} show that the implied volatilities of electricity options depend on the delivery period in a seasonal fashion. Incorporating this idea into a stochastic volatility framework, we start with the following  futures price dynamics under  $\mathbb{Q}$:
 \begin{align}
	df(t,\tau)=s(\tau)\sqrt{\nu(t)}f(t,\tau)dW^f(t)\;,\\
	d\nu(t)=\kappa(\theta-\nu(t))dt+\sigma\sqrt{\nu(t)}dW^\sigma(t)\;.
\end{align}
Here, $s(\tau)$ models the seasonal dependence on the delivery in $\tau$.  Deriving the swap price model as in Section \ref{sec:averaging}, 
the swap price volatility is given by 
\begin{align}
\Sigma(t,\tau_1,\tau_2)=S_1(\tau_1,\tau_2)\sqrt{\nu(t)}\;.
\end{align}
Moreover, the swap's pricing measure  $\widetilde{\mathbb{Q}}$  is defined via the market price of risk
\begin{align}\label{eq:xi_FS}
b_1(t,\tau_1,\tau_2)=- S_2(\tau_1, \tau_2) \sqrt{\nu(t)}\;,
\end{align}
where the constant volatility part is given by
\begin{align}
S_1(\tau_1, \tau_2) =\begin{cases}
\frac{1}{\tau_2-\tau_1}\int_{\tau_1}^{\tau_2}s(u)du\;, & \hat{w}(u)=1\;,\\
\frac{-r}{e^{-r(\tau_2-\tau_1)}-1}\int_{\tau_1}^{\tau_2}e^{-r(u-\tau_1)}s(u)du\;, & \hat{w}(u)=e^{-ru}\;,
\end{cases}
\end{align}
and the constant market price component is characterized by
\begin{align} \label{eq:S_1_2}
S_2(\tau_1, \tau_2) =\begin{cases}
\frac12 \left(  \frac{\int_{\tau_1}^{\tau_2}s^2(u)du}{\int_{\tau_1}^{\tau_2}s(u)du} -S_1(\tau_1,\tau_2)\right)\;, & \hat{w}(u)=1\;,\\
\frac12\left(\frac{\int_{\tau_1}^{\tau_2}e^{-ru}s^2(u)du}{\int_{\tau_1}^{\tau_2}e^{-ru}s(u)du}-S_1(\tau_1,\tau_2)\right)\;, & \hat{w}(u)=e^{-ru}\;.
\end{cases} 
\end{align} Here, $S_1(\tau_1, \tau_2)$ describes the average seasonality in the volatility during the delivery period, and $S_2(\tau_1, \tau_2)$ the relative trade-off between the average squared seasonality (resulting from the average variance of a stream of futures) and
 the squared average seasonality (e.g. the variance part of the average seasonality). 
The swap price dynamics under  $\widetilde{\mathbb{Q}}$ then follow
\begin{align}
dF(t,\tau_1,\tau_2)=&S_1(\tau_1,\tau_2) \sqrt{\nu(t)}F(t,\tau_1,\tau_2)d\widetilde{W}^f(t)\;, \\
d\nu(t) =& \left(\kappa\theta-[\kappa-\sigma\rho S_2(\tau_1,\tau_2)]\nu(t)\right)+\sigma\sqrt{\nu(t)}d\widetilde{W}^{\sigma}(t)\;.
\end{align}
A possible choice for the seasonality function is  \mbox{$s(\tau)=a+b \cos(2\pi(\tau+c))$,} where $a>b>0$ and $c\in[0,1)$ to ensure that the volatility stays positive. In particular, the parameter $a$ represents the long-term variance of the swap seasonally adjusted by the cosine-function driven by the amplitude $b$ and shifted by the parameter $c$, where $c=0$ represents January 1. In this case,  Novikov's condition is satisfied if $\kappa^2> a^2\sigma^2$, such that  the measure change is well defined and $F$ is indeed a true martingale under $\widetilde{\mathbb{Q}}$ (see Appendix \ref{app:Novikov_TS}).
In the setting of Remark \ref{re:1}, we have
\begin{align*}
\Sigma(t,\tau_1,\tau_2)&= \mathbb{E}_U[s(U)]\sqrt{\nu(t)}\;,\\
b_1(t,\tau_1,\tau_2)
&=-\frac12 \frac{ \mathbb{V}_U[s(U)]}{\mathbb{E}_U[s(U)]}\sqrt{\nu(t)}\;,
\end{align*}
for a uniformly distributed random variable $U\sim \mathcal{U}[\tau_1, \tau_2]$.
Having option pricing in view, we would like to mention that  we again preserve the  affine structure of the model, that is as $(\log(f(\cdot,\tau)), \nu(\cdot))$ is affine in the volatility, so is $(\log(F(\cdot,\tau_1, \tau_2)), \nu(\cdot))$ after applying the averaging procedure of Section \ref{sec:averaging}.
\begin{figure}[tb]
\centering
\begin{subfigure}[b]{0.3\textwidth}
\centering
\caption{Volatility Components}
\includegraphics[height=3.8cm]{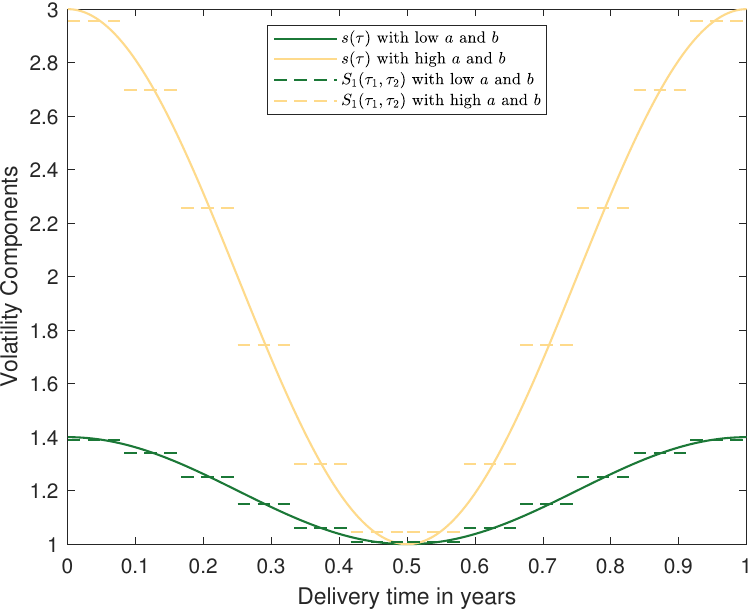}
\end{subfigure}
\hfill
\begin{subfigure}[b]{0.3\textwidth}
\centering
\caption{Market Price Components}
\includegraphics[height=3.8cm]{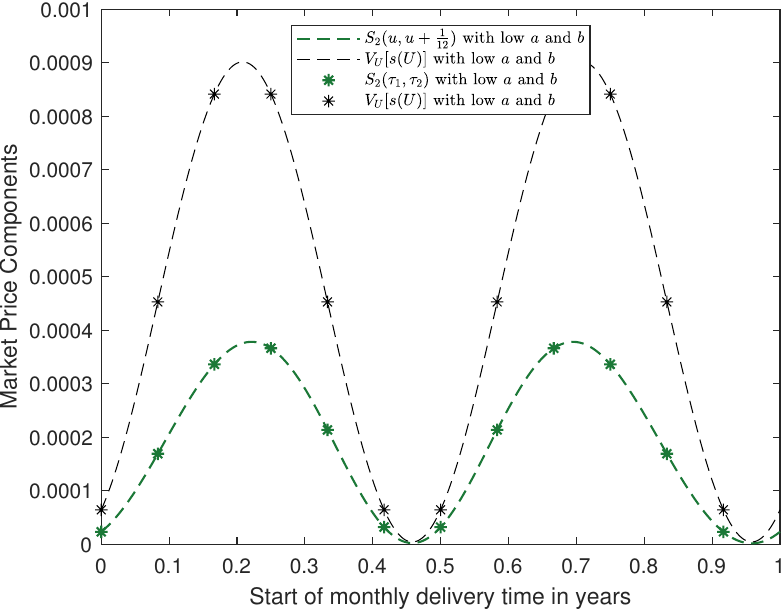}
\end{subfigure}
\hfill
\begin{subfigure}[b]{0.3\textwidth}
	\centering
	\caption{Market Price Components}
	\includegraphics[height=3.8cm]{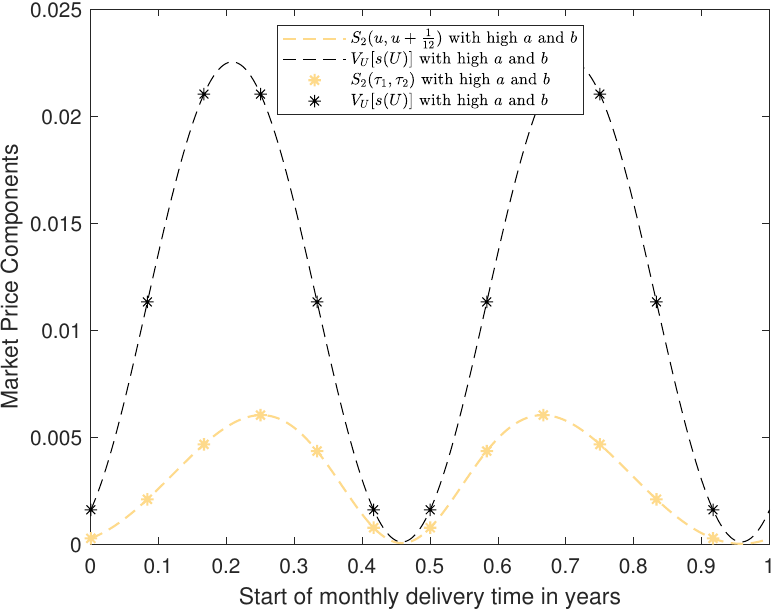}
\end{subfigure}
\caption{Seasonality in the delivery time under constant settlement. Subfigure (a) illustrates $s(\tau)$ for $\tau\in [0,1]$ and $S_1(\tau_1,\tau_2)$ for different monthly delivery times over one year. Subfigure (b) shows $S_2(\tau_1,\tau_2)$ and  the variance of $s(\tau)$ with respect to the start of the delivery times during one year. Note that the asterisk indicates the beginning of a month. 
Subfigure (c) illustrates the same plots as in Subfigure (b) but with higher parameter values for the seasonality function $s(\tau)$.
The parameters can be found in Tables \ref{tab:1} and \ref{tab:1_FanelliSchmeck}.}
\label{fi:Vola_Swap_FS}
\end{figure}
In Figure \ref{fi:Vola_Swap_FS}, the deterministic part of the swap's volatility $S_1(\tau_1,\tau_2)$ is plotted with respect to the delivery time as well as the deterministic part of the market price of risk $S_2(\tau_1,\tau_2)$.
The parameters can be found in Tables \ref{tab:1} and \ref{tab:1_FanelliSchmeck}.
We consider a time horizon of a year including twelve delivery months.
 
In Figure \ref{fi:Vola_Swap_FS} (a), each dashed line represents the swap's variance for the corresponding monthly delivery while the continuous line stands for the continuous seasonality function evolving over the year. Thus each dashed constant is the average seasonality during the corresponding delivery period.
In particular, $S_1$  is the highest in the winter (i.e. $S_1(0,\frac{1}{12})$ and $S_1(\frac{11}{12},1)$) and the lowest in the summer (i.e. $S_1(\frac{5}{12},\frac{1}{2})$ and $S_1(\frac{1}{2},\frac{7}{12})$). 
This phenomenon arises through the chosen parameter $c=0$. Hence, the peak appears directly at the beginning of the year.
In Figures \ref{fi:Vola_Swap_FS} (b) and \ref{fi:Vola_Swap_FS} (c), we consider the constant market price component $S_2$ and the variance of seasonality for lower and higher parameter values $(a,b)$, respectively. In both plots, the variance of seasonality has two peaks in April and October when the changes in $s$ are the largest (by choice of the seasonality function and parameters). Hence, spring and autumn lead to highest variances in the swap. Therefore, the market price component $S_2$ is highest as well driven by the variance component. 
Moreover, we observe that higher volatility values lead to a higher variance in $s$ and to a higher market price component $S_2$.

\subsection{Comparison of the Models} \label{sec:SimComp}
For our simulation study, we applied the Euler-Maruyama procedure to the swap process and the drift-implicit Milstein procedure to the volatility process. The parameters used in this section are summarized in Tables \ref{tab:1} and \ref{tab:Comparison}. 
We would like to comment on the choice of parameter values in the following. We consider a representative delivery in October, i.e.\,$[\tau_1,\tau_2]=[\frac{9}{12}, \frac{10}{12}]$, because it turns out that the market price of delivery risk is highest in October for seasonality in the delivery time. The interest rate is chosen rather small due to the current interest level. 
Corresponding to approximate starting values from recent price series, the initial swap price is assumed to be 50.
For stochastic volatility parameters $\rho, \kappa, \sigma, \alpha, \beta, \gamma$, we have chosen approximate values from the seasonal stochastic volatility model for natural gas by \cite{Arismendi2016}.
For reasons of comparability, we assume that the initial volatility and the constant mean-reversion level coincide with the long-term variance level of the stochastic volatility, i.e.\, $\nu_0=\theta=\alpha$, such that the stochastic volatility by \cite{Arismendi2016} moves around the long-term mean of the remaining models. The range of multiple seasonality parameters $\beta$ is also comparable with estimated variables in \cite{schneider2018samuelson} for the exponential sinusoidal model in agricultural futures markets. Whenever $\beta=0$, we consider the benchmark case corresponding to the Heston model.
Note that the variables for the models in Sections \ref{sec:TS} and \ref{sec:FS} are summarized in Tables \ref{tab:1_Samuelson} and \ref{tab:1_FanelliSchmeck} and are inspired by estimates from \cite{schneider2018samuelson}, \cite{kiesel} and \cite{fanelli2019seasonality}. Note that the variables in Table \ref{tab:1_Samuelson} are adapted for better comparability as \cite{schneider2018samuelson} consider agricultural products which may differ in the size of parameters. The fitted Samuelson parameter in a deterministic volatility setting for electricity swaps investigated by \cite{kiesel} is much higher with $\lambda\approx 1.4$. 
Finally, in Table~\ref{tab:Comparison} the most pronounced values from Tables \ref{tab:1_Arismendi}--\ref{tab:1_FanelliSchmeck} are chosen for the comparison of the three models and the benchmark.
For each model, they fulfill the Feller condition to ensure that the stochastic volatility stays strictly positive as well as the Novikov condition such that the measure change is well defined (see Section \ref{sec:Arismendi}--\ref{sec:FS} for details).
Note that the initial swap price volatility $\Sigma(0,\tau_1,\tau_2)$ might be different for each model even if the initial value $\nu_0$ of the Cox-Ingersoll-Ross process is always the same.
\begin{table}[tb]
	\centering
	\begin{tabular}{C{13mm} C{13mm} C{13mm} C{13mm} C{13mm} C{13mm} C{13mm} C{13mm} C{13mm}}
		\midrule
		\multicolumn{9}{c}{Joint Parameters} \\
		\midrule\addlinespace
		$f_0$ & $\nu_0$ & $\tau_1$ & $\tau_2$ & $\rho$ & $r$ & $\kappa$ & $\sigma$ & $\theta$ \\ \addlinespace
		$50$ & $0.3$ & $0.75$ & $\tau_1+\frac{1}{12}$ & $0.4$ & $0.005$ & $3$ & $0.6$ & $0.3$ \\\addlinespace
		\bottomrule
	\end{tabular}%
	\caption{Parameters for the simulations fixed for all models.}
	\label{tab:1}%
\end{table}%
\begin{table}[tb]
\centering
\begin{minipage}{0.32\textwidth}
\begin{tabular}{C{10mm} C{24mm} C{10mm}}
\midrule
\multicolumn{3}{c}{Seasonality in Trading Days} \\    \addlinespace
\midrule
$\alpha$ & $\beta$ & $\gamma$ \\ \addlinespace
$0.3$  & $[0,~ 0.5,~ 0.75]$ & $0.5$ \\ \addlinespace
\bottomrule
\end{tabular}%
\caption{Parameters:\\ Seasonality in trading days.}
\label{tab:1_Arismendi}%
\end{minipage}
\hfill
\begin{minipage}{0.2\textwidth}
\centering
\begin{tabular}{C{1mm} C{25mm} C{1mm}}
\midrule
\multicolumn{3}{c}{Samuelson} \\  \addlinespace
\midrule
& $ \lambda$ & \\ \addlinespace
&  $[0.25,~ 1,~ 3]$    & \\ \addlinespace
\bottomrule
\end{tabular}%
\caption{Parameters:\\ Samuelson effect.}
\label{tab:1_Samuelson}%
\end{minipage}
\hfill
\begin{minipage}{0.33\textwidth}
\centering
\begin{tabular}{C{15mm} C{15mm} C{10mm}}
\midrule
\multicolumn{3}{c}{Seasonality in the Delivery} \\ \addlinespace
\midrule
$a$     & $b$     & $c$ \\ \addlinespace
$[1.2,~ 2]$     & $[0.2,~ 1]$   & $0$ \\ \addlinespace
\bottomrule
\end{tabular}%
\caption{Parameters:\\ Seasonality in delivery periods.}
\label{tab:1_FanelliSchmeck}%
\end{minipage}
\end{table}%
\begin{table}[tb]
\centering
\begin{tabular}{C{13mm} C{15mm} C{13mm} C{13mm} C{13mm} C{13mm} C{13mm} 
	}
\midrule
\multicolumn{7}{c}{Parameters for Comparisons} \\
\midrule\addlinespace
$\alpha$ & $\beta$ & $\gamma$ & $\lambda$ & $a$ & $b$ & $c$ 
\\ \addlinespace
$0.3$ & $[0, 0.75]$ & $0.5$ & $3$ & $2$ & $1$ & $0$  
\\\addlinespace
\bottomrule
\end{tabular}%
\caption{Parameters for the comparisons of all models, where $\beta=0$ refers to the benchmark case and $\beta=0.75$ to the model with seasonality in the trading day.}
\label{tab:Comparison}%
\end{table}%

\begin{figure}[tb]
\centering
\begin{subfigure}[b]{0.3\textwidth}
\centering
\caption{Swap Price Evolution}
\includegraphics[height=3.8cm]{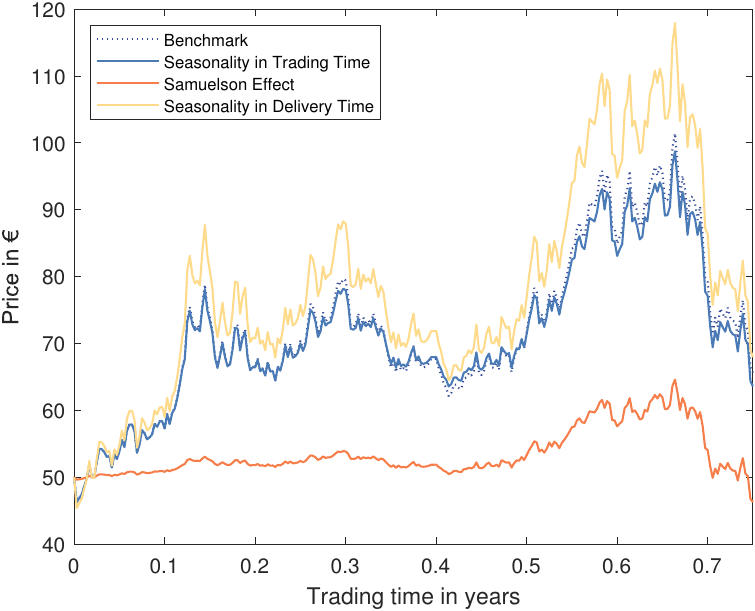}
\end{subfigure}
\hfill
\begin{subfigure}[b]{0.3\textwidth}
	\centering
	\caption{Swap Price Volatility}
	\includegraphics[height=3.8cm]{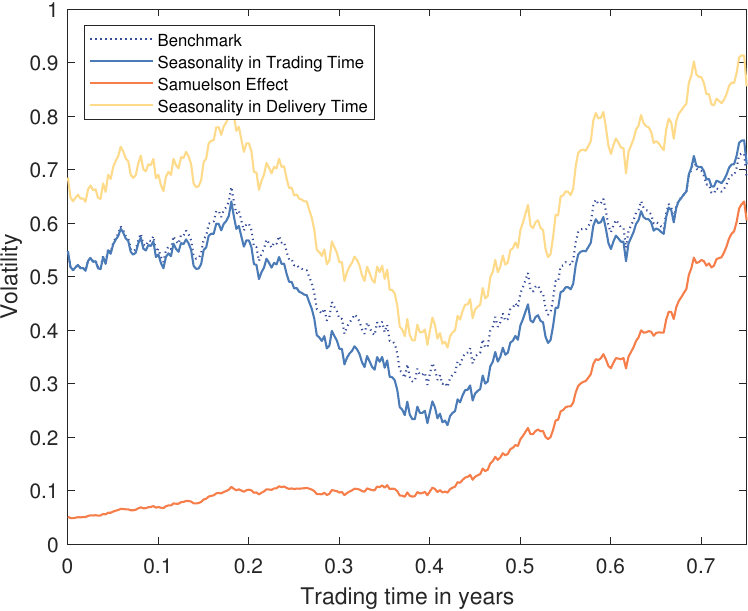}
\end{subfigure}
\hfill
\begin{subfigure}[b]{0.3\textwidth}
\centering
\caption{Market Prices Delivery Risk}
\includegraphics[height=3.8cm]{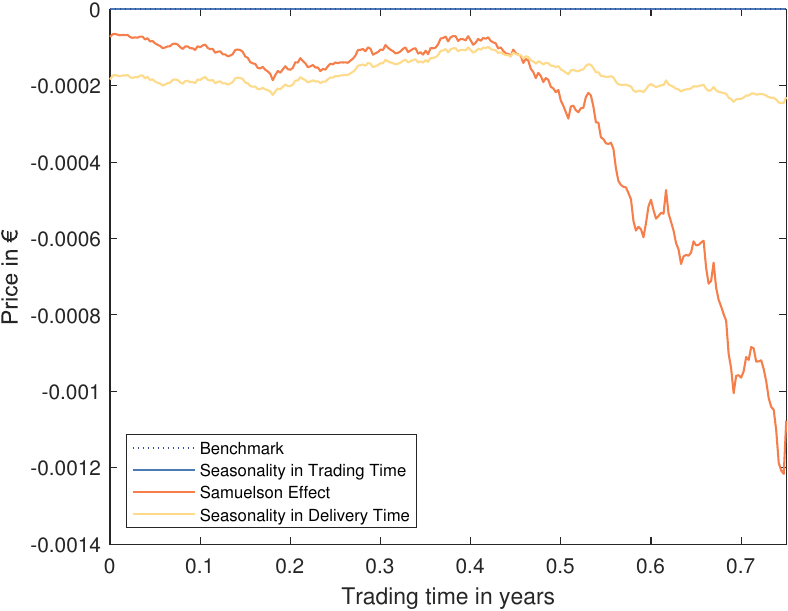}	
\end{subfigure}
\caption{Comparison of the models. Subfigure (a): Swap prices $F(\cdot,\tau_1,\tau_2)$ for each example. 
	Subfigure (b): The swap price volatility $\Sigma(\cdot,\tau_1,\tau_2)$ for each example.
	Subfigure (c): The market prices $b_1(\cdot,\tau_1,\tau_2)$ for each example.
	The trajectories are based on the parameters in Tables \ref{tab:1} and \ref{tab:Comparison}, where the benchmark is based on $\beta=0$. }
\label{fi:Vola_Swaps_Comparison_QQ}
\end{figure}

In Figure \ref{fi:Vola_Swaps_Comparison_QQ}, we have plotted the evolution of the swap prices and its stochastic volatility as well as the market prices of delivery risk of all three considered models and the benchmark  under the measure $\widetilde{\mathbb{Q}}$. For better comparison, we use the same Brownian increments for each model.
Our time scale reflects nine months starting from January with delivery in October.
In the orange trajectory, we can clearly observe the Samuelson effect, which diminishes the stochastic volatility at the beginning and pushes in the end towards $d_1(\frac{1}{12})\sqrt{\nu(\tau_1)}$.
Moreover, the swap price volatility with seasonality in the trading day is oscillating around the benchmark swap price volatility. 
The swap price volatility with seasonality in the delivery time moves above, but in parallel to, the benchmark.
This is caused by the choice of parameters since $S_1(\tau_1,\tau_2)> 1$. The market price of delivery risk is zero for the delivery independent models, whereas the market price oscillates around $-0.0002$ driven by seasonalities in the delivery time and goes to $-0.0012$ when the Samuelson effect is considered.

The dependence on the delivery time in electricity swap markets is significant, based on the existing empirical validation (see \cite{kiesel}, \cite{fanelli2019seasonality}).
In
our opinion, the larger the proportion of renewable energy in an economy, the stronger the role of seasonality in the delivery period, and the more important will be the adequate implementation.  
A combination with term-structure effects might be even more precise but the combination of both effects still has to be confirmed empirically. 
In real market situations, we recommend to
implement at least seasonality in the delivery time thus leading to a market price of delivery risk being non-zero.
Our concept of the market price of risk for delivery periods therefore enables more accurate pricing.

\section{Further Arbitrage Considerations} \label{sec:Arbitrage_Cons}
So far, we have considered a market with one swap contract.
Nevertheless, in electricity markets, typically more than one swap is  traded at the same time. For example, at the EEX, the next 9 months, 11 quarters, and 6 years are available.  In Section \ref{sec:averaging_multi}, we address the issue of arbitrage in a market consisting of $N$ monthly delivering swaps and then discuss  a market with overlapping delivery periods in Section \ref{sec:averaging_overlapping}.

\subsection{Absence of Arbitrage in a Market with $N$ Swaps} \label{sec:averaging_multi}
In this section, we consider a market with $N$ swap contracts having subsequent monthly delivery periods $(\tau_m,\tau_{m+1}]$   for $m=1,\ldots,N$. 
According to the First Fundamental Theorem of Asset Pricing, the market is arbitrage-free if there exists 
a measure $\widetilde{\mathbb{Q}}$ under which all swap dynamics are martingales.
In a market with $N$ assets,  $N$ Brownian motions are needed such that a market price of risk exists (see, e.g., \cite{Shreve2004}). Therefore, we add another factor for each contract
and the underlying futures price dynamics are given by
\begin{equation} \label{eq:futurespricedynamicsmultitime}
df(t,\tau) = f(t,\tau) \sum_{j=1}^{N}\sigma_j(t,\tau)dW_j^{f}(t)\;, \quad \quad f(0,\tau) = f_0 >0\;,
\end{equation}  
where $W_j^F$, for $j=1,\ldots, N$, are independent standard Brownian motions under $\mathbb{Q}$.
As in Section \ref{sec:averaging}, we define the swap price with delivery period $(\tau_m,\tau_{m+1}]$, for $m=1,\ldots,N$ via geometric averaging 
\begin{align*}
F(t, \tau_m, \tau_{m+1}) :=& \exp\left(\int_{\tau_m}^{\tau_{m+1}}w(u, \tau_m, \tau_{m+1})\log(f(t,u))du\right)\;. 
\end{align*}
The resulting swap price dynamics for the monthly delivery period  $(\tau_m,\tau_{m+1}]$, $m=1,\ldots,N,$ are given by
\begin{equation}\label{eq:FunderQ_multi}
\begin{split}
\frac{d F(t, \tau_m, \tau_{m+1})}{F(t, \tau_m, \tau_{m+1})} =& -\frac{1}{2}\sum_{j=1}^{N}\left(\int_{\tau_m}^{\tau_{m+1}}w(u, \tau_m, \tau_{m+1})\sigma_j^2(t,u)du -\Sigma_j^2(t,\tau_m, \tau_{m+1}) \right)dt\\&+\sum_{j=1}^{N}\Sigma_j(t,\tau_m, \tau_{m+1})~dW_j^{f}(t)\;,
\end{split}
\end{equation}
where $\Sigma_j(t,\tau_m, \tau_{m+1}):=\int_{\tau_m}^{\tau_{m+1}}w(u, \tau_m, \tau_{m+1})\sigma_j(t,u)du$ analogous to Equation \eqref{eq:Sigma}.
Then the standard theory for multidimensional markets (see, e.g., \cite{Shreve2004}) leads to the market price of risk equations and a risk-neutral probability measure.


\subsection{Absence of Arbitrage in a Market with Overlapping Swaps} \label{sec:averaging_overlapping}
\begin{figure}
	\begin{tikzpicture}
	\node[fill = gray!40, shape = rectangle, rounded corners,
	minimum width = 16cm, font = \sffamily] (A) at (0,0){Year} ;
	\node[fill = gray!40, shape = rectangle, rounded corners,
	minimum width = 1cm, font = \sffamily] (C) at (-7.25,-2)  {Month 1} ;
	\node[fill = gray!40, shape = rectangle, rounded corners,
	minimum width = 1cm, font = \sffamily] (D) at (-5.5,-2)  {Month 2} ;
	\node[fill = gray!40, shape = rectangle, rounded corners,
	minimum width = 1cm, font = \sffamily] (E) at (-3.75,-2)  {Month 3} ;
	\node[fill = gray!40, shape = rectangle, rounded corners,
	minimum width = 3cm, font = \sffamily] (F) at (-0.5,-2)  {Quarter 2} ;
	\node[fill = gray!40, shape = rectangle, rounded corners,
	minimum width = 3cm, font = \sffamily] (G) at (3,-2)  {Quarter 3} ;
	\node[fill = gray!40, shape = rectangle, rounded corners,
	minimum width = 3cm, font = \sffamily] (H) at (6.5,-2)  {Quarter 4} ;
	\node[fill = gray!40, shape = rectangle, rounded corners,
	minimum width = 1cm, font = \sffamily] (J) at (-3,-4)  {Month 4} ;
	\node[fill = gray!40, shape = rectangle, rounded corners,
	minimum width = 1cm, font = \sffamily] (K) at (-0.5,-4)  {Month 5} ;
	\node[fill = gray!40, shape = rectangle, rounded corners,
	minimum width = 1cm, font = \sffamily] (L) at (2,-4)  {Month 6} ;
	\path[draw,->] 
	(A.south) -- (C.north);
	\path[draw,->] 
	(A.south) -- (D.north);
	\path[draw,->] 
	(A.south) -- (E.north);
	\path[draw,->] 
	(A.south) -- (F.north);
	\path[draw,->] 
	(A.south) -- (G.north);
	\path[draw,->] 
	(A.south) -- (H.north);
	\path[draw,->] 
	(F.south) -- (J.north);
	\path[draw,->] 
	(F.south) -- (K.north);
	\path[draw,->] 
	(F.south) -- (L.north);
	\end{tikzpicture}
	\caption{The cascading procedure of overlapping electricity swap contracts.}
	\label{fig:cascading}
\end{figure}
In electricity markets, it is possible to trade into overlapping delivery periods.
For example, the swap contract on the next quarter of the year is available as well as the three swaps on the corresponding months (e.g., the first quarter of the year and monthly contracts for January to March).
Also here, arbitrage has to be excluded: It should not matter if the electricity is bought via a quarterly contract or the corresponding three underlying monthly contracts.

One has to find a pricing measure under which all swap processes, the monthly and the quarterly ones, are martingales.
If we would price an overlapping contract using the geometric averaging procedure, 
we would have that
\begin{align*}
F^{overl}(t,\tau_1,\tau_{N+1})=&~e^{\int_{\tau_1}^{\tau_{N+1}}w(u,\tau_1,\tau_{N+1})\log(f(t,u))du}
=\prod_{m=1}^{N}F(t,\tau_m,\tau_{m+1})^{w_m}\;,
\end{align*}
where $w_m=\frac{\int_{\tau_m}^{\tau_{m+1}}\hat{w}(u)du}{\int_{\tau_1}^{\tau_{N+1}}\hat{w}(u)du}$. The price of 
 the quarterly swap would be the product of the monthly contracts. This might create arbitrage opportunities:
In general, the product of martingales is not a martingale anymore.
In this framework, the so-called cascading process of overlapping contracts offers a solution.
The cascading process  describes the division of an overlapping contract into its building blocks. A swap contract delivering over a quarter is transformed into its  corresponding monthly swap contracts at its maturity (see Figure \ref{fig:cascading}).
Analogously, the price of a yearly swap contract is converted into the first 3~monthly contracts and the subsequent 3 quarterly contracts.
Each quarterly contract will be cascaded later.
The monthly contracts thus play the role of building blocks for overlapping contracts and are also called atomic contracts.
Consequently, the quarterly and yearly swap contracts can be seen derivatives on the monthly contracts, and we propose to price them as such.
In particular,
\begin{align*}
F^{overl}(t,\tau_1,\tau_{N+1}):=\sum_{m=1}^{N}w_m F(t,\tau_m,\tau_{m+1}),
\end{align*}
where $N=12$ for a yearly contract and $N=3$ for a quarter swap.
If we have found a pricing measure under which all atomic swap price processes are martingales, then the process $F^{overl}$ is also a martingale since the sum of $\widetilde{\mathbb{Q}}$-martingales stays a martingale under $\widetilde{\mathbb{Q}}$.

\section{Electricity Options} \label{sec:options}
We consider a European option with strike price $K>0$ and exercise time $T\leq\tau_1$ written on an electricity swap contract delivering in $(\tau_1,\tau_2]$. 
In Section \ref{sec:averaging}, we have determined an equivalent measure $\widetilde{\mathbb{Q}}$, such that 
the swap price process $F(\cdot,\tau_1,\tau_2)$ is a martingale. Hence,  $\widetilde{\mathbb{Q}}$ can be used as pricing measure for derivatives on the swap. 
In general, $\widetilde{\mathbb{Q}}$ depends on $\tau_1$ and $\tau_2$ since it includes a risk premium for the delivery period as discussed in Remark \ref{re:1}.
Hence, the pricing measure is tailor-made for this particular contract.

\subsection{An Application of the Heston-Methodology}
Motivated by the market models considered in Section \ref{sec:Trafo_Models}, we 
stick to a general factorizing volatility structure $\Sigma(t,\tau_1,\tau_2)=S(t,\tau_1,\tau_2)\sqrt{\nu(t)}$, where
\begin{align} \label{eq:Vola_general}
S(t,\tau_1,\tau_2)=\mathbb{E}\left[s(t,U)\right]
\end{align}
identifies averaged seasonalities and term-structure effects for a random variable $U$ with density $w(u,\tau_1,\tau_2)$ (see also Remark \ref{re:1}). 
We assume that $s(t,u)$ is positive and bounded by $R$, 
so that the swap price dynamics 
 \begin{align}
	dF(t,\tau_1,\tau_2)=& S(t,\tau_1,\tau_2)\sqrt{\nu(t)}F(t,\tau_1,\tau_2)d\widetilde{W}^f(t)\;, \label{eq:F_Q_tilde}\\
	d\nu(t)=&\left(\kappa\theta(t)-(\kappa-\sigma\rho\xi(t,\tau_1,\tau_2))\nu(t)\right)dt+\sigma\sqrt{\nu(t)}d\widetilde{W}^\sigma(t)\;.
\end{align}
is a  $\widetilde{\mathbb{Q}}$-martingale if $2\kappa^2>\sigma^2R^2$ (see Appendix \ref{app:Novikov_TS}). 
The market price of delivery risk is given 
 by $b_1(t,\tau_1,\tau_2) =-\xi(t,\tau_1,\tau_2)\sqrt{\nu(t)}$, where
\begin{align} \label{eq:Xi_general}
	\xi(t,\tau_1,\tau_2)=\frac{1}{2}\frac{\mathbb{V}_U\left[s(t,U)\right]}{\mathbb{E}_U\left[s(t,U)\right]}\;.
\end{align}
The price of the corresponding electricity call and put options at time $t\in[0,T]$ is given by the risk-neutral valuation formula
\begin{align}\label{eq:riskneutralvaluation}
&C(t, \tau_1,\tau_2)=\mathbb{E}_{\tilde{\mathbb{Q}}}\left[e^{-r(T-t)} \left(F(T,\tau_1, \tau_2) -K \right)^+ | \mathcal{F}_t\right]\;,\\
&P(t, \tau_1,\tau_2)=\mathbb{E}_{\tilde{\mathbb{Q}}}\left[e^{-r(T-t)} \left(K-F(T,\tau_1, \tau_2) \right)^+ | \mathcal{F}_t\right]\;,
\end{align}
see, e.g., \cite{Shreve2004}. 
The dynamics of the logarithmic swap price $X(t):=\log(F(t,\tau_1,\tau_2))$ are given by 
\begin{align}
dX(t)= -\frac{1}{2} S^2(t,\tau_1,\tau_2)\nu(t)dt+S(t,\tau_1,\tau_2)\sqrt{\nu(t)}d\widetilde{W}^f(t)\;. \label{eq:logSwap}
\end{align}
We skip the dependencies of the log-price on the delivery period for notational convenience.
We thus have the following result for call options:
\begin{theorem} \label{theorem:OptionPrice}
Let $x:=\log F$ denote the logarithmic swap price observed at time $t\leq T$.
Then, the electricity call option price at time $t\leq T$ with strike price $K>0$ is given by
\begin{align} 
C(t,\tau_1,\tau_2)=e^{-r(T-t)}\left( e^{x}~(1-Q_1(t, x, \nu; \log(K)))-K~(1-Q_2(t, x, \nu; \log(K))) \right)\;,\label{eq:option_price}
\end{align}
where the probabilities of exercising the call option at time $t$ defined in \eqref{eq:ProbOfExercising1} and \eqref{eq:ProbOfExercising2} are given by
\begin{align} 
1-Q_k(t, x, \nu; \log(K))=\frac{1}{2}+ \frac{1}{\pi} \int_{0}^{\infty}Re\left(\frac{e^{-i \phi \log(K)} \hat{Q}_k(t,x, \nu;\phi)}{i\phi}\right)d\phi\;, \quad k=1,2\;. \label{eq:probability}
\end{align}
The characteristic functions  $\hat{Q}_k(t,x,\nu;\phi)$ are given by
\begin{align} \label{eq:Q_hat}
\hat{Q}_k(t,x,\nu;\phi)= e^{\Psi_{0k}(t,T,\phi)+\nu\Psi_{1k}(t,T,\phi)+i \phi x}\;, \quad\quad k=1,2\;,
\end{align}
where $\Psi_{0k}(t,T,\phi)$ and $\Psi_{1k}(t,T,\phi)$ for $k=1,2$ solve the following system of differential equations
\begin{align}
\frac{\partial \Psi_{1k}}{\partial t} =&
-\frac{1}{2} \sigma^2 \Psi^2_{1k}+ \left(\beta_k(t,\tau_1,\tau_2)-\rho \sigma S(t,\tau_1,\tau_2) i \phi \right)\Psi_{1k}+\left(\frac{1}{2} \phi^2-\alpha_k i\phi
\right) S^2(t,\tau_1,\tau_2)\;, \label{eq:Riccati} \\
\frac{\partial \Psi_{0k}}{\partial t}=&
-\Psi_{1k}\kappa\theta(t)\;, \label{eq:ODE2} 
\end{align}
subject to $\Psi_{0k}(T,T,\phi)=\Psi_{1k}(T,T,\phi)=0$ and 
for $\alpha_1=\frac12$, $\alpha_2=-\frac12$, $\beta_1(t,\tau_1,\tau_2)=\kappa-\sigma\rho(\xi(t,\tau_1,\tau_2)+S(t,\tau_1,\tau_2))$,  $\beta_2(t,\tau_1,\tau_2)=\kappa-\sigma\rho\xi(t,\tau_1,\tau_2)$.
\end{theorem}
The proof follows the Heston procedure and can be found in Appendix \ref{app:Proof}.
There exists a unique solution to each 
Riccati equation for $k=1,2$ (see Appendix \ref{app:Ricatti}) and thus also for $\Psi_{01}$ and $\Psi_{02}$. Then, the characteristic functions in \eqref{eq:Q_hat} are uniquely determined.
The related put option price can be determined by the \textit{Put-Call-Parity}. In particular,
\begin{align}
P(t,\tau_1,\tau_2)=C(t,\tau_1,\tau_2)-e^{-r(T-t)}\left(F(t,\tau_1,\tau_2)-K\right) \;.
\end{align}

\begin{remark}
In Theorem \ref{theorem:OptionPrice}, we denote the probability of exercising the option using $Q_k$ as a cumulative probability defined in \eqref{eq:ProbOfExercising1} and \eqref{eq:ProbOfExercising2}. Note, that \cite{Heston} is based on the probability of finishing in the money  $P_k$, that is $Pr[X^{t,x,\nu}\geq log(K)]=1-Pr[X^{t,x,\nu}<\log(K)]$.
\end{remark}

The value of this result depends strongly on the tractability of the Riccati equations \eqref{eq:Riccati} for $k=1,2$.
In the classical Heston model, all coefficients of the Riccati equations \eqref{eq:Riccati} for $k=1,2$ are constant, so that one can find an analytical solution. This phenomenon can also be observed in the setting with seasonalities in the delivery time as the coefficients are independent of trading time. Note, that our findings are an extension of the model by \cite{fanelli2019seasonality} since they focus on deterministic volatility only.
For time-dependent coefficients, it is not clear that an analytical solution or closed-form expression exists. 
In \cite{Arismendi2016}, the mean reversion level $\theta(t)$ of the stochastic volatility process is seasonal, but 
as $\theta(t)$  does not appear in the Riccati equations and in \eqref{eq:ODE2} only, an analytical solution exists.
\cite{schneider2018samuelson} include the Samuelson effect  such that the futures price dynamics have time-dependent coefficients.
Nevertheless, the volatility process has constant coefficients.
The Samuelson term appears in the Riccati equations and \cite{schneider2018samuelson} are able to give a solution depending on Kummer functions. In our framework, the Samuelson effect appears in the drift of the stochastic volatility via the market price of delivery risk, making the Riccati equations more complicated (see Section \ref{sec:option_TS}).

\subsection{The Effect of Seasonalities and Samuelson on the Swaps' Riccati Equation} \label{sec:examples}
In this section, we state the differential equations \eqref{eq:Riccati} and \eqref{eq:ODE2} for each model, and discuss how to solve them.
Furthermore, we compare the corresponding call option prices numerically, for which we choose the end of the option contracting period as $T=\tau_1=0.75$.
Let us mention, that for the classical Riccati equation setting,
which corresponds to the Heston model with constant coefficients, the
exact formula for the solution can be found easily. In general, the equation has to
be solved numerically. We discuss this situation in the subsequent sections.

\subsubsection{Seasonal Dependence on the Trading Day} \label{sec:option_A}
In the setting of  \eqref{eq:Trafo_Arismendi_F} and \eqref{eq:Trafo_Arismendi_nu}, option pricing has been treated by \cite{Arismendi2016}. 
Recall, that $\Psi_{0k}(t,T,\phi)$ and $\Psi_{1k}(t,T,\phi)$ solve the following system of differential equations 
\begin{align*}
	\frac{d \Psi_{1k}}{dt}(t,T,\phi)=&-\frac{1}{2}\sigma^2\Psi_{1k}^2(t,T,\phi)+\left(\beta_k-\rho\sigma i \phi \right)\Psi_{1k}(t,T,\phi)+\frac{1}{2}\phi^2-\alpha_k i \phi\;,\\
\frac{d \Psi_{0k}}{dt}(t,T,\phi) =&- \kappa \theta(t) \Psi_{1k}(t,T,\phi)\;,
\end{align*}
for $\alpha_1=\frac{1}{2}$, $\alpha_2=-\frac{1}{2}$, $\beta_1=\kappa-\sigma\rho$, and $\beta_2=\kappa$.
Since all coefficients of the Riccati equations for $k=1,2$ (the first equation of the system) are constant, the solutions can be calculated by
\begin{align*}
\Psi_{1k}(t,T,\phi)=\frac{1}{\sigma^2}\left(\beta_k-\sigma\rho\phi i-\delta_k\right)\frac{1-e^{-\delta_k(T-t)}}{1-g_ke^{-\delta_k(T-t)}}\;, \quad k=1,2\;,	
\end{align*}
using the little Heston trap procedure by \cite{Albrecher2007},
where 
\begin{align*}
\delta_k:=&\sqrt{(\beta_k-\sigma\rho\phi i)^2+\sigma^2(\phi^2-2\alpha_k\phi i)}\;, \quad \text{and } \quad
g_k:=\frac{\beta_k-\sigma\rho\phi i -\delta_k}{\beta_k-\sigma\rho\phi i+\delta_k}\;.
\end{align*}
Finally, numerical integration leads to the solution of $\Psi_{01}(t,T,\phi)$ and $\Psi_{02}(t,T,\phi)$ (see Section \ref{sec:numerics}).

\begin{figure}[tb]
	\centering
	\begin{subfigure}[b]{0.3\textwidth}
		\centering
		\caption{Seasonality in Trading Day}
		\includegraphics[height=3.8cm]{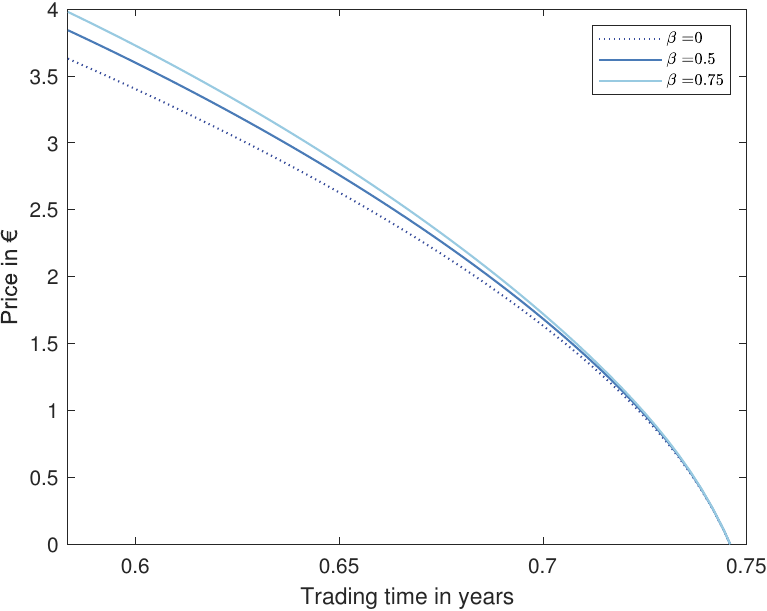}
	\end{subfigure}
	\hfill
	\begin{subfigure}[b]{0.3\textwidth}
		\centering
		\caption{Samuelson Effect}
		\includegraphics[height=3.8cm]{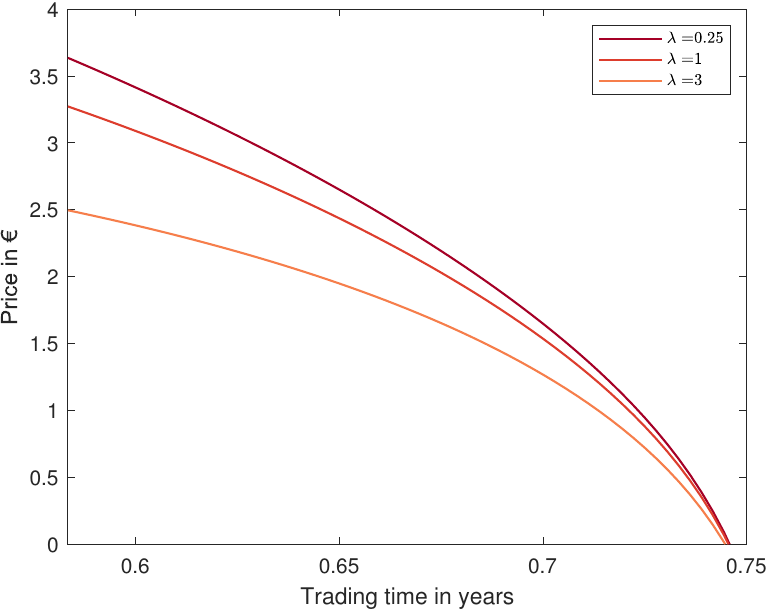}
	\end{subfigure}
	\hfill
	\begin{subfigure}[b]{0.3\textwidth}
		\centering
		\caption{Seasonality in Delivery}
		\includegraphics[height=3.8cm]{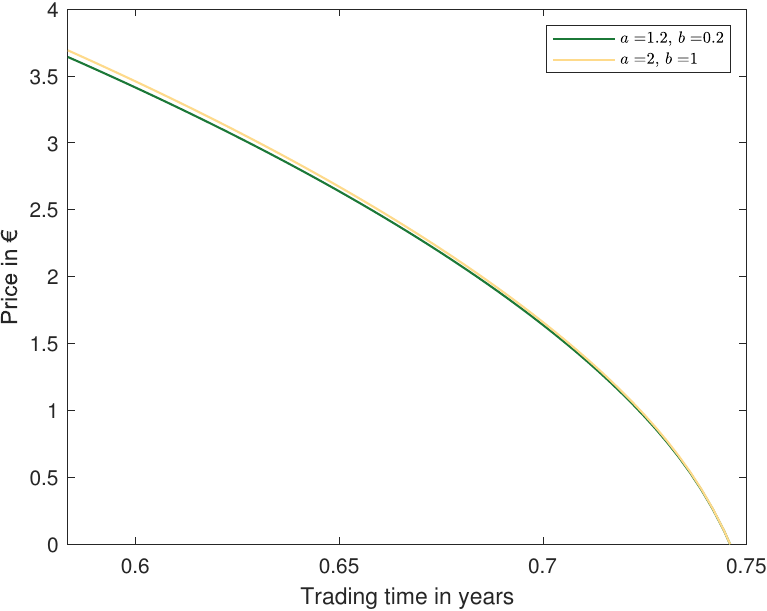}	
	\end{subfigure}
\caption{Call option prices over the last two trading months for a fixed strike $K=52$ under constant settlement. Subfigure (a): Call prices with seasonality in the trading day.
Subfigure (b): Call prices with Samuelson effect. Subfigure (c): Call prices with seasonality in the delivery time. The parameters can be found in Tables \ref{tab:1}--\ref{tab:1_FanelliSchmeck}.
}
\label{fi:OptionPrices_Effects_QQ}
\end{figure}

In Figure \ref{fi:OptionPrices_Effects_QQ} (a), we illustrate the call option prices for different amplitudes in the mean-reversion level over the last two trading months based on the parameters in Tables \ref{tab:1} and \ref{tab:1_Arismendi}. The calculations are conducted by using the analytical solution for the Riccati equations. The solutions for $\Psi_{0k}$ are attained using the Runge-Kutta method. 
As proposed by \cite{Arismendi2016}, we apply a trapezoidal integration scheme to obtain the integral values for each strike which are mandatory to determine the corresponding probabilities $1-Q_1$ and $1-Q_2$ as in \eqref{eq:probability} and thus the related call option price for the considered strike at a specific point in time (see \eqref{eq:option_price}).
For a fixed strike price $K=52$, we observe the basic feature that all call option prices are decreasing over time. Furthermore, we observe that the higher the amplitude $\beta$ of the mean-reversion-level of the stochastic volatility, the higher is the call option price.
The closer we reach the expiration date, the smaller is the difference between each call option price based on different amplitudes $\beta$.

\subsubsection{Samuelson Effect} \label{sec:option_TS}
In the setting of Section \ref{sec:TS}, the resulting dynamics under $\widetilde{\mathbb{Q}}$ are given by \eqref{eq:F_TS} and \eqref{eq:nu_TS}. Under the measure $\widetilde{\mathbb{Q}}$, the Samuelson effect appears in the drift term of the stochastic volatility.

$\Psi_{0k}(t,T,\phi)$ and $\Psi_{1k}(t,T,\phi)$ for $k=1,2$ solve the following two systems of differential equations:
\begin{align*}
\frac{d \Psi_{11}}{dt}(t,T,\phi)=&-\frac{1}{2}\sigma^2\Psi_{11}^2(t,T,\phi)+\frac12 d_1(\tau_2-\tau_1)^2 (\phi^2-i\phi)e^{-2\lambda(\tau_1-t)}\\&+\Big(\kappa-\sigma\rho\Big[d_2(\tau_2-\tau_1)+ d_1(\tau_2-\tau_1)(1+i\phi)\Big]e^{-\lambda(\tau_1-t)}\Big)\Psi_{11}(t,T,\phi)\;,\\
\frac{d \Psi_{01}}{dt}(t,T,\phi) =& -\kappa \theta \Psi_{11}(t,T,\phi)\;,
\end{align*}
and
\begin{align*}
\frac{d \Psi_{12}}{dt}(t,T,\phi)=&-\frac{1}{2}\sigma^2\Psi_{12}^2(t,T,\phi)+\frac12 d_1(\tau_2-\tau_1)^2 (\phi^2+i\phi)e^{-2\lambda(\tau_1-t)}\\&+\Big(\kappa-\sigma\rho\Big[ d_2(\tau_2-\tau_1)+i \phi d_1(\tau_2-\tau_1)\Big]e^{-\lambda(\tau_1-t)} \Big)\Psi_{12}(t,T,\phi)\;,\\
\frac{d \Psi_{02}}{dt}(t,T,\phi) =& -\kappa \theta \Psi_{12}(t,T,\phi)\;,
\end{align*}
where $d_1(x)$ and $d_2(x)$ are defined in \eqref{eq:d1_d2}.
Compared to \cite{schneider2018samuelson}, the Samuelson effect appears additionally in front of $\Psi_{11}$ and $\Psi_{12}$.
This leads to the setting of time-dependent coefficients in the Riccati-type equations.
The explicit solution is expressed in terms of hypergeometric expressions.

Figure \ref{fi:OptionPrices_Effects_QQ} (b) illustrates call option prices during the last two trading months based on the default parameters introduced in Tables \ref{tab:1} and \ref{tab:1_Samuelson}.
We use the Runge-Kutta method with adaptive step size to solve both systems of differential equations.
The trapezoidal integration of the integrands with respect to $\phi$ leads to the cumulative distributions $Q_1$ and $Q_2$ for each strike price $K$.
An application of Equation \eqref{eq:option_price} gives the corresponding call option prices for each strike.
In fact, we approximate the analytic expressions for $\Psi_{11}$, $\Psi_{12}$, $\Psi_{01}$, and $\Psi_{02}$ since two coefficients in the Riccati equation include the Samuelson effect.

For models with time-dependent $\theta, \sigma,$ and $\rho$, the method by \cite{Benhamou} can be applied with the help of  a volatility of variance expansion using the Lewis representation.
For time-dependent coefficients of piece-wise constant structure, one can also use the model of
\cite{Nogel}.
However, in general, the solution has to be found numerically.
 Most of them concern the one dimensional case, for example, standard second order finite difference methods, see \cite{Tavella}. More recently, results include stochastic volatility with high-order compact finite difference schemes such as Crank–Nicolson scheme, see \cite{During}.

In order to investigate the impact of the Samuelson effect, we set the parameter $\lambda$ to three different values $0.25$, $1$, and $3$.
We observe higher call option prices for smaller Samuelson parameters, as the underlying swap becomes more volatile.
With increasing time to maturity, the price differences become even larger (see Figure \ref{fi:OptionPrices_Effects_QQ} (b)).  

\subsubsection{Delivery-Dependent Seasonality} \label{sec:option_FS}
Finally, we consider the resulting call option prices corresponding to Section \ref{sec:FS}.
$\Psi_{0k}(t,T,\phi)$ and $\Psi_{1k}(t,T,\phi)$ for $k=1,2$ solve the following two systems of differential equations:
\begin{align*}
\frac{d \Psi_{11}}{dt}(t,T,\phi)=&-\frac{1}{2}\sigma^2\Psi_{11}^2(t,T,\phi)+\Big(\kappa-\sigma\rho\Big[S_2(\tau_1,\tau_2)+ S_1(\tau_1,\tau_2)(1+i \phi)\Big]\Big)\Psi_{11}(t,T,\phi)\\&
+\frac12 S_1(\tau_1,\tau_2)^2 (\phi^2-i\phi)\;,\\
\frac{d \Psi_{01}}{dt}(t,T,\phi) =& -\kappa \theta \Psi_{11}(t,T,\phi)\;,
\end{align*}
and
\begin{align*}
\frac{d \Psi_{12}}{dt}(t,T,\phi)=&-\frac{1}{2}\sigma^2\Psi_{12}^2(t,T,\phi)+\Big(\kappa-\sigma\rho\Big[ S_2(\tau_1,\tau_2)+ i \phi S_1(\tau_1,\tau_2)\Big] \Big)\Psi_{12}(t,T,\phi)\\&+\frac12 S_1(\tau_1,\tau_2)^2 (\phi^2+i\phi)\;,\\
\frac{d \Psi_{02}}{dt}(t,T,\phi) =& -\kappa \theta \Psi_{12}(t,T,\phi)\;.
\end{align*}
The differential equations can be solved analytically, while all coefficients are constant. The solutions are given by
\begin{align*}
\Psi_{0k}(t,T,\phi)=& \frac{\kappa \theta}{\sigma^2} \Bigg[\Big(\beta_k(\tau_1,\tau_2)-\sigma\rho \phi i -\delta_k(\tau_1,\tau_2)\Big)(T-t)-2\log\left(\frac{1-g_k (\tau_1,\tau_2) e^{-\delta_k (\tau_1,\tau_2)(T-t)}}{1-g_k(\tau_1,\tau_2)} \right) \Bigg]\;,\\
\Psi_{1k}(t,T,\phi)=&\frac{1}{\sigma^2}\left(\beta_k(\tau_1,\tau_2)-\sigma\rho\phi i-\delta_k(\tau_1,\tau_2)\right)\frac{1-e^{-\delta_k(\tau_1,\tau_2)(T-t)}}{1-g_k(\tau_1,\tau_2)e^{-\delta_k(\tau_1,\tau_2)(T-t)}}\;,
\end{align*}
where $\beta_1(\tau_1,\tau_2)=\kappa-\sigma\rho(S_2(\tau_1,\tau_2)+S_1(\tau_1,\tau_2))$, $\beta_2(\tau_1,\tau_2)=\kappa-\sigma\rho S_2(\tau_1,\tau_2)$ and
\begin{align*}
&\delta_k(\tau_1,\tau_2):=\sqrt{(\beta_k(\tau_1,\tau_2)-\sigma\rho\phi i)^2+\sigma^2(\phi^2-2\alpha_k\phi i)}\;,\\
&g_k(\tau_1,\tau_2):=\frac{\beta_k(\tau_1,\tau_2)-\sigma\rho\phi i -\delta_k(\tau_1,\tau_2)}{\beta_k(\tau_1,\tau_2)-\sigma\rho\phi i+\delta_k(\tau_1,\tau_2)}\;.
\end{align*}
The delivery dependent seasonality model is able to incorporate delivery dependent effects, while being highly tractable and fast to implement.
In Figure \ref{fi:OptionPrices_Effects_QQ} (c), we visualize the call option prices over the last two trading months based on the parameters in Tables \ref{tab:1} and \ref{tab:1_FanelliSchmeck}.
For the calculations, we use the analytical solutions for $\Psi_{01}, \Psi_{02}, \Psi_{11}$, and $\Psi_{12}$.
As before, numerical integration leads to the desired call option price for each considered strike. 
We observe that the call option prices slightly increase for higher swap volatility values $(a,b)$ for $K=52$.

\subsubsection{Numerical Comparison of the Effects} \label{sec:numerics}
\begin{figure}[tb]
	\centering
	\begin{subfigure}[b]{0.45\textwidth}
		\centering
		\caption{Integrand for $k=1$}
		\includegraphics[height=5.3cm]{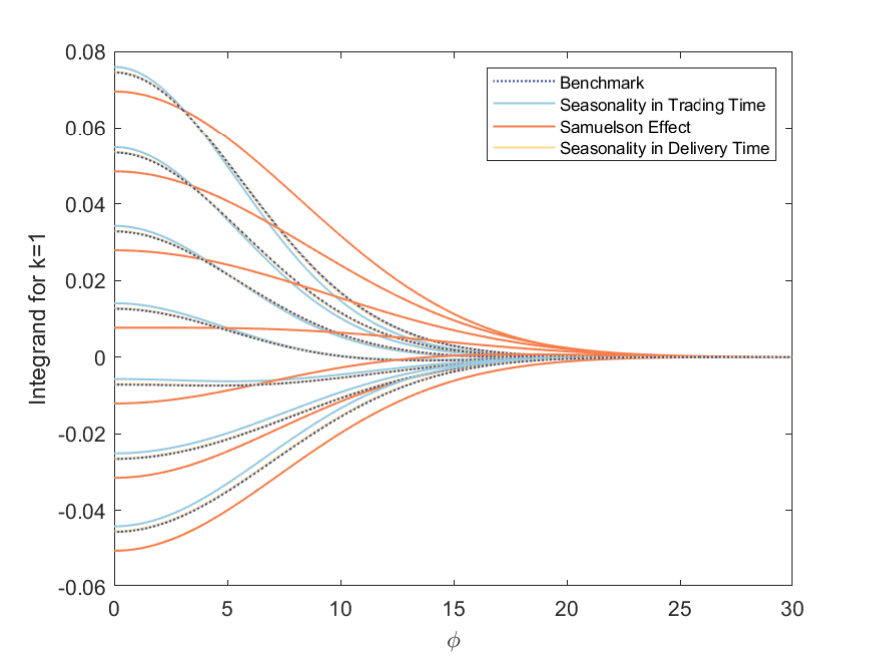}
	\end{subfigure}
	\hfill
	\begin{subfigure}[b]{0.45\textwidth}
		\centering
		\caption{Integrand for $k=2$}
		\includegraphics[height=5.3cm]{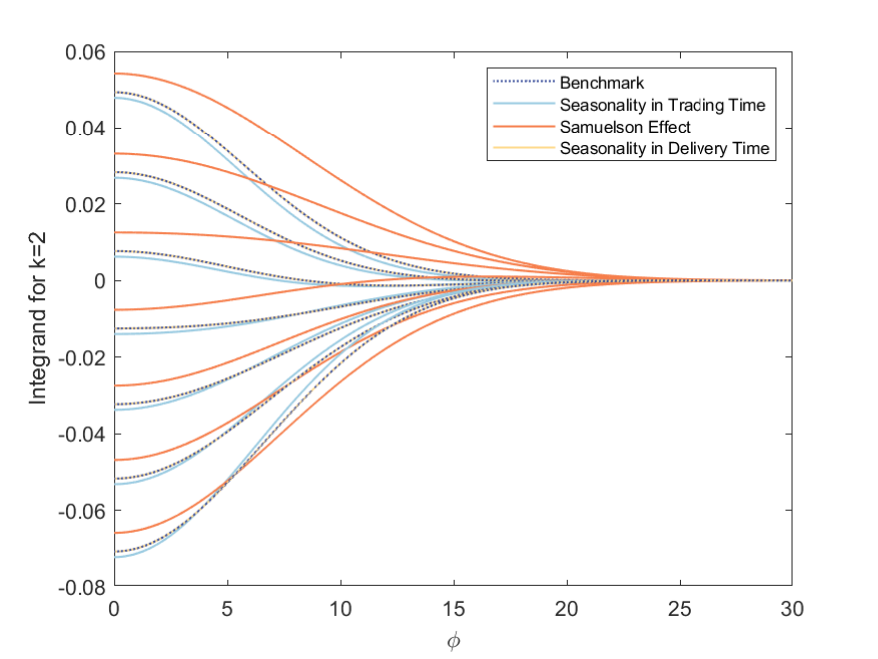}	
	\end{subfigure}
	\caption{Integrands of Equation \eqref{eq:probability} evaluated one month before maturity for all strikes $K=47, \ldots,53$.  Subfigure (a): Integrand for $k=1$ for each model. Subfigure (b): Integrand for $k=2$ for each model. The parameters can be found in Tables \ref{tab:1} and \ref{tab:Comparison}, where the benchmark is based on  $\beta=0$. }
	\label{fi:Integrands_comparison_10_PS7}
\end{figure}

\begin{figure}[tb]
	\centering
	\begin{subfigure}[b]{0.45\textwidth}
		\centering
		\caption{Option Prices over Strikes}
		\includegraphics[height=5.3cm]{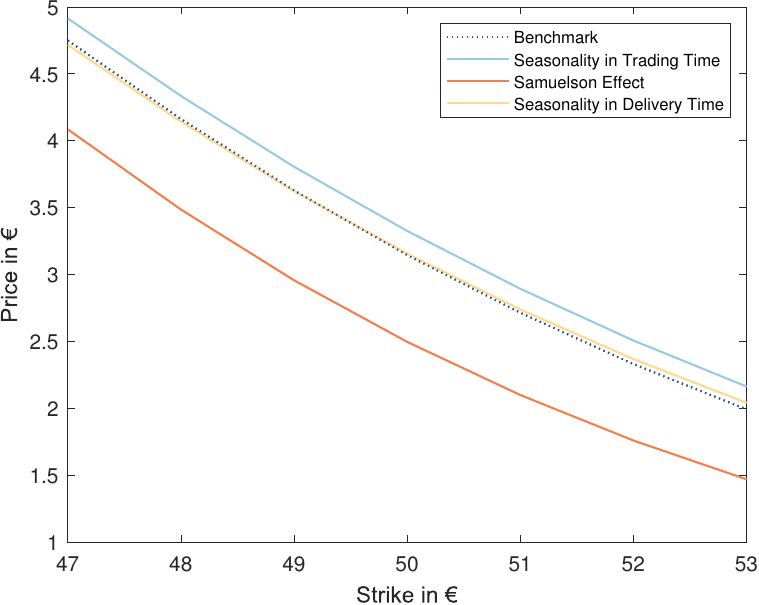}
	\end{subfigure}
	\hfill
	\begin{subfigure}[b]{0.45\textwidth}
		\centering
		\caption{Option Prices over Time}
		\includegraphics[height=5.3cm]{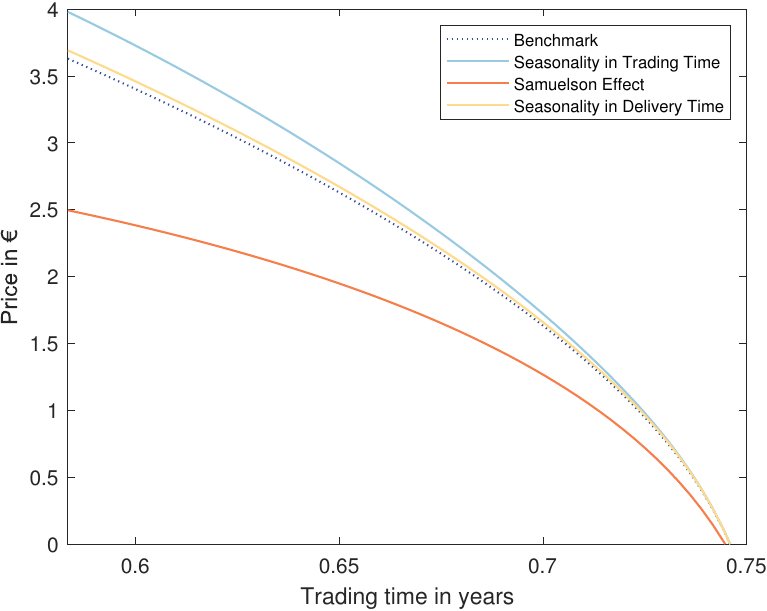}	
	\end{subfigure}
	\caption{Subfigure (a): Option prices for each model one month before maturity. Subfigure (b): Option prices for each model over the last two trading months for a fixed strike $K=52$. The parameters can be found in Tables \ref{tab:1} and \ref{tab:Comparison}, where the benchmark refers to  $\beta=0$. }
	\label{fi:OptionPrices_IV_Comparison_QQ}
\end{figure}

In this section, we focus on concrete numerical examples based on the transformed models in Section \ref{sec:Trafo_Models} and Section \ref{sec:examples}.
We consider the integrands for each model as well as the resulting call option prices.
For comparative reasons, we have chosen the same parameters for all electricity swap price models (see Tables \ref{tab:1} and \ref{tab:Comparison}).
In order to determine both integrands of \eqref{eq:probability} for each model, we calculate the solution to the system of ordinary differential equations as in Section \ref{sec:option_A}--\ref{sec:option_FS}.
We used the analytical solution of the Riccati equations with seasonality in the trading day and in the delivery time.
We get  a certain integrand depending on $\phi$ for each strike price.
The possible oscillation of the integrand can have a negative influence on the numerical procedure since the standard quadrature can fail, see
\cite{Rouah}. 
We can observe that both integrands are relatively smooth for each model and converge to zero around $\phi\approx30$ (see Figure \ref{fi:Integrands_comparison_10_PS7}).
For the integration, we apply the standard trapezoidal rule truncating the integrand boundaries by $\phi\in[1e-7,30]$ due to the converting behavior.
Plugging the integral value into \eqref{eq:probability} leads to the call option prices \eqref{eq:option_price} (see Figure~\ref{fi:OptionPrices_IV_Comparison_QQ}).

To compare the resulting prices, we plot the call option prices based on the parameters in Tables \ref{tab:1} and \ref{tab:Comparison}
one month before maturity for seven strikes $K$ ranging from $47,\ldots,53$ (see Figure  \ref{fi:OptionPrices_IV_Comparison_QQ} (a)).
One month before maturity, the Samuelson effect leads to lowest call prices whereas seasonality in the trading day leads to highest call option prices. The benchmark and seasonality in the delivery period have values in between. 
For in the money strikes, the benchmark leads to higher call option prices than seasonality in the delivery period and vice versa for out of the money strikes. 

In Figure  \ref{fi:OptionPrices_IV_Comparison_QQ} (b), we consider the price evolution
starting two months before the swap contract expires for a fixed strike price $K=52$. 
During the last two trading months seasonality in the trading time leads to highest call option prices and the Samuelson effect to lowest call prices. 
For $K=52$, the seasonality in the delivery period is smaller than seasonality in the trading time but higher than the benchmark. The ordering between the Samuelson effect, the benchmark, and seasonality in the delivery time arises through the fact that $S_1(\tau_1,\tau_2)>1>d_1(\frac{1}{12})$.

\section{Conclusion} \label{sec:summary}
We suggest the use of a pricing framework for swaps and options in electricity markets.
Moreover, we introduced an equivalent martingale measure for the swap that explicitly depends on its delivery period and can be used to price electricity options.
Geometric averaging on the delivery period is the key element here.
The market price of delivery risk for an individual contract is specified by the trade-off  
between the variance of the swap on the one hand and the weighted average variance of a stream of futures on the other hand.
We considered futures price models from the recent literature and provide the corresponding swap price models.
Moreover, we investigated the effect of 
 seasonal dependence on the trading day, the Samuelson effect, and delivery-dependent seasonality in line with
\cite{Arismendi2016}, \cite{schneider2018samuelson}, and \cite{fanelli2019seasonality}, respectively.
Whenever the futures and thus the swap price volatility are independent of the delivery time, the market price of delivery risk is zero.
On the other hand, typically observed characteristics of the electricity market, such as seasonalities in the delivery and term-structure effects, instead impact the market price of the delivery risk.

Moreover, we provided an outlook of our model in the case of several atomic and overlapping contracts. For each additional atomic contract, new uncertainty occurs, and  a further Brownian motion is thus needed within the futures price.
The pricing procedure can be applied as before.
Overlapping contracts are treated as derivatives of the underlying atomic swap contracts, which is justified by the cascading process (see Figure \ref{fig:cascading}).

All examples are characterized by a volatility structure in the spirit of the Heston model. The affine model structure of the futures is inherited by the swaps, thereby leading us to follow the Heston methodology for option pricing.
We investigated the call option price for seasonal dependence on the trading day, the Samuelson effect, and delivery-dependent seasonality. 
Whenever the deterministic volatility part is independent of the trading time, the corresponding Riccati equations can be solved analytically.
For the Samuelson effect, the deterministic part of the volatility is time-dependent, and we showed that a unique solution exists. Furthermore, we provided a numerical method to solve the Riccati equations.
It might be very interesting to investigate the market price of risk for delivery periods empirically for the delivery dependent models presented in this work as well as for a combination of seasonal and term-structure effects. However, these considerations will be a question for future research.

In conclusion, this paper treats each  electricity swap as a proper contract on the market and suggests a pricing measure that is tailor-made for this particular contract, which includes acknowledging the existence of the delivery period. Our pricing framework allows for the evaluation of option prices in line with the Heston method.

\newpage
\appendix
\section*{\appendixname} 
\numberwithin{equation}{subsection}
\setcounter{equation}{0}
\renewcommand*{\thesubsection}{\Alph{subsection}} 
\subsection{Technical Requirements} \label{app:tech_requ}
\begin{enumerate}
\item For the model \eqref{eq:futurespricedynamicsonetime} and \eqref{eq:futurespricedynamicsonetime2}, we make the following assumtions:
\begin{enumerate}
\item  The conditions by Yamada and Watanabe need to be satisfied (see also  \cite{KaratzasShreve1991}; Proposition 2.13). In particular, we assume
\begin{align*}
&a(t,\tau,\sigma)\colon [0,\tau_1]\times(\tau_1,\tau_2]\times\mathbb{R}^+\to \mathbb{R}\;,\\
&c(t,\tau,\sigma)\colon [0,\tau_1]\times(\tau_1,\tau_2]\times\mathbb{R}^+\to \mathbb{R}\;,
\end{align*}
are Borel-measurable functions and $\sigma^2 = \{\sigma^2(t,\tau)~|~0\leq t\leq \tau\leq\tau_2 \}$ is a  stochastic process with continuous sample paths.
Further, we assume
\begin{itemize}
	\item $\vert a(t,\tau,x) -a(t,\tau,y)\vert \leq K \vert x-y \vert$ for some positive constant $K>0$ with $x,y\in\mathbb{R}^+$,
	\item $\vert c(t,\tau,x) -c(t,\tau,y)\vert \leq H(\vert x-y \vert)$ for $x,y\in\mathbb{R}^+$ where $H\colon [0,\infty)\to  [0,\infty)$ is an increasing function with $H(0)=0$ and $\int_{(0,\epsilon)} H^{-2}(u)du=\infty, ~\forall \epsilon >0$,
\end{itemize}
which guarantees that there exists a unique strong solution for \eqref{eq:futurespricedynamicsonetime2}. In particular, $\sigma^2(t,\tau)$ is adapted to the filtration $\mathcal{F}_t$.
\item Next, we assume that $f=\{f(t,\tau)~|~0\leq t \leq \tau \leq \tau_2\}$ is a stochastic process with continous sample paths. It directly follows that $\sigma(t,\tau)f(t,\tau)$ is process Lipschitz and thus functional Lipschitz. 
Then, by \cite{Protter2005} (see Theorem 7; p. 253) Equation \eqref{eq:futurespricedynamicsonetime} admits a unique strong solution.

\item In order to attain that \eqref{eq:futurespricedynamicsonetime} is a $\mathbb{Q}$-martingale, we assume that the Novikov condition  (see, e.g., \cite{KaratzasShreve1991}; Proposition 5.12) is satisfied, that is
\begin{align}
\mathbb{E}_\mathbb{Q}\left[e^{\frac12 \int_{0}^{\tau}\sigma^2(t,\tau)dt}\right]<\infty\;.
\end{align}
\end{enumerate}
\item For the geometric weightening approach \eqref{eq:GeomAverage} we need to apply the stochastic Fubini Theorem (see \cite{Protter2005}; Theorem 65; Chapter IV. 6). Therefore, we assume that 
\begin{itemize}
\item $(t, u,\omega)\to w(u,\tau_1,\tau_2)\sigma(t,u)$ is jointly progressivly measurable, 
\item $\mathbb{E}_\mathbb{Q}\left[\int_{0}^{\tau_1}\int_{\tau_1}^{\tau_2}w^2(u,\tau_1,\tau_2)\sigma^2(t,u)du~dt\right]<\infty\;$.
\end{itemize}
\end{enumerate}

\subsection{An Application of Girsanov's Theorem for the Examples}
\label{app:Novikov_TS}
We want to check if Novikov's condition is satisfied, that is  $\mathbb{E}_\mathbb{Q}\left[e^{\frac12 \int_{0}^{\tau_1}b^2_1(t,\tau_1,\tau_2)dt}\right]<\infty$ (see, e.g., \cite{KaratzasShreve1991}). 

In the case of \cite{schneider2018samuelson}, we can find specific upper and lower boundaries for the deterministic part since
$e^{-\lambda(\tau_1-t)}\in[0,1]$ and $d_2(\tau_2-\tau_1)\in[-\frac12 \frac{1}{\lambda(\tau_2-\tau_1)},\frac12]$.
Hence, 
\begin{align}
\mathbb{E}_\mathbb{Q}\left[e^{\frac12 \int_{0}^{\tau_1}b^2_1(t,\tau_1,\tau_2)dt}\right]
=\mathbb{E}_\mathbb{Q}\left[e^{\frac12 \int_{0}^{\tau_1}d_2(\tau_2-\tau_1)^2e^{-2\lambda(\tau_1-t)}\nu(t)dt}\right]
\leq \mathbb{E}_\mathbb{Q}\left[e^{-\tilde{u} \int_{0}^{\tau_1}\nu(t)dt}\right]\;,
\end{align}
where $\tilde{u}:=-\frac12 d_2(\tau_2-\tau_1)^2$. Following \cite{cont_tankov2003financial} (see Section 15.1.2) there exists an explicit, finite expression for the last expectation if  $\kappa^2+2\sigma^2 \tilde{u}>0$.

In the case of \cite{fanelli2019seasonality}, we can again find specific upper and lower boundaries for \eqref{eq:xi_FS} since $s(u)=a+b\cos(2\pi(c+u))\in[0,2a]$ for $a>b>0$ and thus $s^2(u)\leq 2a~s(u)$. In particular, $S_2(\tau_1,\tau_2)\in[-a,a]$.
Hence,
\begin{align}
\mathbb{E}_\mathbb{Q}\left[e^{\frac12 \int_{0}^{\tau_1}b^2_1(t,\tau_1,\tau_2)dt}\right]
=\mathbb{E}_\mathbb{Q}\left[e^{\frac12 \int_{0}^{\tau_1}S_2(\tau_1,\tau_2)^2\nu(t)dt}\right]
\leq \mathbb{E}_\mathbb{Q}\left[e^{-\tilde{u}  \int_{0}^{\tau_1}\nu(t)dt}\right]\;,
\end{align}
where $\tilde{u}:=- \frac12 a^2$. As before, the last expectation is limited if  $\kappa^2+2\sigma^2\tilde{u}>0$, i.e. $\kappa^2>a^2\sigma^2$.

In the general case of Section \ref{sec:options},  we assume that $s(t,u)$ is positive and bounded.
As $s(t,u)\in[\epsilon,R]$, for $\epsilon\in(0,R)$ small and 
$\xi(t,\tau_1,\tau_2)\in[-\frac12 \frac{R^2}{\epsilon},\frac12 \frac{R^2}{\epsilon}]$.
Hence, 
\begin{align}
\mathbb{E}_\mathbb{Q}\left[e^{\frac12 \int_{0}^{\tau_1}b^2_1(t,\tau_1,\tau_2)dt}\right]
=\mathbb{E}_\mathbb{Q}\left[e^{\frac12 \int_{0}^{\tau_1}\xi(t,\tau_1,\tau_2)^2\nu(t)dt}\right]
\leq \mathbb{E}_\mathbb{Q}\left[e^{-\tilde{u} \int_{0}^{\tau_1}\nu(t)dt}\right]\;,
\end{align}
where $\tilde{u}:=-\frac14 \frac{R^4}{\epsilon^2}$. As before, if $\kappa^2+2\sigma^2\tilde{u}>0$, that is if  $2\kappa^2> \frac{R^4}{\epsilon^2} \sigma^2$,  then Novikov's condition is satisfied. 
This result holds for a general weight function type $\hat{w}$.
\subsection{Proof of Theorem \ref{theorem:OptionPrice}} \label{app:Proof}
\begin{proof}
We can write 
\begin{align}
C(t,\tau_1,\tau_2,) = e^{-r(T-t)}\mathbb{E}_{\widetilde{\mathbb{Q}}} \left[e^{X_t}\mathds{1}_{X_T\geq\log(K)} \vert \mathcal{F}_t\right] - e^{-r(T-t)} K~ \mathbb{E}_{\widetilde{\mathbb{Q}}} \left[\mathds{1}_{X_T\geq\log(K)} \vert \mathcal{F}_t\right]\;.  
\end{align}
Due to the Markovian structure, an application of the \textit{Independence Lemma} (see, e.g., \cite{Shreve2004}; cf. Lemma 2.3.4) leads to
\begin{align*}
 C(t,\tau_1,\tau_2)=c_1(t,X(t),\nu(t))-c_2(t,X(t),\nu(t))\;,
\end{align*}
where
\begin{align}
&c_1(t,x,\nu) = e^{-r(T-t)}e^{x}~(1-Q_1(t, x, \nu; \log(K)))\;, \label{eq:C_1}\\
&c_2(t,x,\nu) = e^{-r(T-t)}K~(1-Q_2(t, x, \nu; \log(K)))\;, \label{eq:C_2} \\
	&Q_1(t, x, \nu; \log(K)):=\tilde{\tilde{\mathbb{Q}}}\left[X^{t,x,\nu}(T)< \log(K)\right]\;, \label{eq:ProbOfExercising1}\\	
	&Q_2(t, x, \nu; \log(K)):=\tilde{\mathbb{Q}}\left[X^{t,x,\nu}(T)< \log(K)\right]\;, \label{eq:ProbOfExercising2}
\end{align}
where the probability measure $\widetilde{\widetilde{\mathbb{Q}}}$ is defined by $\frac{d\widetilde{\widetilde{\mathbb{Q}}}}{d\widetilde{\mathbb{Q}}}=e^{ -\frac{1}{2}\int_{0}^{T}S^2(t, \tau_1, \tau_{2}) \nu(t) dt+\int_{0}^{T} S(t, \tau_1, \tau_{2})\sqrt{\nu(t)} dW^{f}(t)}$.

For $k=1,2$, $e^{-rt}c_k(t,X(t),\nu(t))$  are martingales under $\widetilde{\mathbb{Q}}$.
Hence, $c_k(t, x,\nu)$ solves
\begin{equation} \label{eq:PDE_1}
\frac{\partial c_k(t, x, \nu)}{\partial t} + (\mathcal{A}_t c_k)(t, x, \nu) = rc_k(t, x, \nu), ~~\text{for } k=1,2,
\end{equation}
subject to the terminal conditions $c_1(T,x,\nu)=e^x \mathds{1}_{x\geq\log(K)}$ and $c_2(T,x,\nu)=K \mathds{1}_{x\geq\log(K)}$,
by an application of the discounted Feynman Kac Theorem (see, e.g., \cite{Shreve2004}; cf. Theorem 6.4.3 and Ch. 6.6).
For a function $f$ depending on $x$ and $\nu$, the generator of $(X, \nu)$ is given by
\begin{equation} \label{eq:generator1}
\begin{split}
(\mathcal{A}_t f)(x,\nu)
=& -\frac{1}{2} \frac{\partial f}{\partial x}S^2(t,\tau_1,\tau_2)\nu
+ \frac{\partial f}{\partial \nu} [\kappa \theta(t)-(\kappa-\sigma\rho\xi(t,\tau_1,\tau_2))\nu]\\
&+\frac{1}{2}\frac{\partial^2 f}{(\partial x)^2}S^2(t,\tau_1,\tau_2)\nu
+\frac{1}{2}\frac{\partial^2 f}{(\partial \nu)^2}\sigma^2 \nu
+ \frac{\partial^2 f}{\partial x \partial \nu}\rho \sigma S(t,\tau_1,\tau_2)\nu\;.
\end{split}
\end{equation} 
If we plug  \eqref{eq:C_1} and \eqref{eq:C_2} inside the partial differential equation (PDE) \eqref{eq:PDE_1}, we end up with 
\begin{equation}\label{eq:PDE_main}
\begin{split}
&	
\frac{\partial Q_k}{\partial t} +\alpha_k S^2(t,\tau_1,\tau_2)\nu \frac{\partial Q_k}{\partial x}
+\left(\kappa \theta(t)-\beta_k(t,\tau_1,\tau_2)\nu\right)\frac{\partial Q_k}{\partial \nu}\\
&+\frac12 S^2(t,\tau_1,\tau_2)\nu\frac{\partial^2 Q_k}{(\partial x)^2}
+\frac12\sigma^2\nu \frac{\partial^2 Q_k}{(\partial \nu)^2}+\rho\sigma\nu S(t,\tau_1,\tau_2)\frac{\partial^2 Q_k}{\partial x \partial \nu}
= 0\;,
\end{split}
\end{equation}
for $\alpha_1=\frac12$, $\alpha_2=-\frac12$, $\beta_1(t,\tau_1,\tau_2)=\kappa-\sigma\rho(\xi(t,\tau_1,\tau_2)+S(t,\tau_1,\tau_2))$, and $\beta_2(t,\tau_1,\tau_2)=\kappa-\sigma\rho\xi(t,\tau_1,\tau_2)$.
This PDE can be solved by a martingale depending on the solutions of the  dynamics 
\begin{align*}
dX_k(t)=\alpha_k S^2(t,\tau_1,\tau_2)\nu_k(t) dt+S(t,\tau_1,\tau_2)\sqrt{\nu_k(t)} d\widetilde{W}^f(t)\;,\\
d\nu_k(t)=\left(\kappa \theta(t)-\beta_k(t,\tau_1,\tau_2)\nu_k(t)\right)dt+\sigma \sqrt{\nu_k(t)} d\widetilde{W}^\sigma(t)\;.
\end{align*}
Following Heston, the corresponding characteristic function solves \eqref{eq:PDE_main} as well.
Note, that the underlying model structure is of affine type since the PDE is linear in $\nu$. 
The characteristic function is thus of exponential affine form (see \cite{Duffie_DAP}):
\begin{align}
\hat{Q}_k(t,x,\nu;\phi)= \mathbb{E}_{Q_k}\left[e^{i\phi X_k^{t,x,\nu}(T)}\right] = e^{\Psi_{0k}(t,T,\phi)+\nu\Psi_{1k}(t,T,\phi)+i \phi x}\;, \quad\quad k=1,2\;,\label{eq:expaffinefunction1}
\end{align}
for $\phi\in\mathbb{R}$,
 where $\Psi_{0k}\colon [0,T]\times[0,\tau_1)\times \mathbb{R}\to \mathbb{C}$ and $\Psi_{1k}\colon  [0,T]\times [0,\tau_1)\times \mathbb{R}\to \mathbb{C}$ are time-dependent functions satisfying $\Psi_{0k}(T,T,\phi)=0$ and $\Psi_{1k}(T,T,\phi)=0$ at terminal time $T$.
The last term in \eqref{eq:expaffinefunction1} is added in order to ensure the terminal condition 
\begin{align}
\hat{Q}_k(T,x,\nu;\phi) = e^{i \phi x}\;.
\end{align}
For notational convenience, we drop the time and space indices such that $\Psi_{0k}:=\Psi_{0k}(t,T,\phi)$, $\Psi_{1k}:=~\Psi_{1k}(t,T,\phi)$, and $\hat{Q}_k:=\hat{Q}_k(t,x,\nu;\phi)$.   
Plugging \eqref{eq:expaffinefunction1} into the PDEs of \eqref{eq:PDE_main} for $k=1,2$ and rearranging terms yields
\begin{align*}
\hat{Q}_k\Bigg[
\nu \Big[ &
\frac{\partial \Psi_{1k}}{\partial t}
+\alpha_k S^2(t,\tau_1,\tau_2)i\phi
-\Psi_{1k}\beta_k(t,\tau_1,\tau_2)
-\frac{1}{2} S^2(t,\tau_1,\tau_2) \phi^2
+\frac{1}{2} \sigma^2 \Psi^2_{1k}\\
&+\rho \sigma S(t,\tau_1,\tau_2) i \phi \Psi_{1k}
\Big]
+\frac{\partial \Psi_{0k}}{\partial t}
+\Psi_{1k}\kappa\theta(t)
\Bigg] =0\;.
\end{align*}
Since $\hat{Q}_k>0$ for $k=1,2$ and $\nu>0$ by definition, we apply the \textit{separation of variables argument} (see \cite{Duffie_DAP}; cf. p. 150) to achieve the following differential equations
\begin{align*}
\frac{\partial \Psi_{1k}}{\partial t}
=-\frac{1}{2} \sigma^2 \Psi^2_{1k}
+ \left(\beta_k(t,\tau_1,\tau_2)-\rho \sigma S(t,\tau_1,\tau_2) i \phi \right)\Psi_{1k}
+\left(\frac{1}{2} \phi^2-\alpha_k i\phi\right) S^2(t,\tau_1,\tau_2)
\end{align*}
of Riccati-type and
$\frac{\partial \Psi_{0k}}{\partial t}=-
\Psi_{1k}\kappa\theta(t)\;,$
subject to $\Psi_{0k}(T,T,\phi)=0$ and $\Psi_{1k}(T,T,\phi)=0$ for $k=1,2$.

An application of the Fourier inversion technique (see \cite{Gil}) to  \eqref{eq:expaffinefunction1} leads to the cumulative distribution functions $Q_1$ and $Q_2$ given by
\begin{align}
Q_k(t,x,\nu; \log(K))=\frac{1}{2}- \frac{1}{\pi} \int_{0}^{\infty}Re\left(\frac{e^{-i \phi \log(K)} \hat{Q}_k(t,x, \nu;\phi)}{i\phi}\right)d\phi\;, \quad k=1,2\;. \label{eq:integrand}
\end{align}
\end{proof}

\subsection{On the Solutions of the Differential Equations} \label{app:Ricatti}
To show that there exists a unique solution for the Riccati equations $\Psi_{1k}(t,T,\phi)$ for $k=1,2$, transfer them to a homogenoeus second order linear differential equation using the substitution  $\Psi_{1k}(t,T,\phi)=\frac{z_k'(t,T)}{\frac12\sigma^2~z_k(t,T)}$ (see, e.g., \cite{Poljanin}). Rewrite the resulting differential equation as a system of first order equations in line with \cite{Walter1996}(cf. p. 103f.).
Then, Theorem VI (see \cite{Walter1996}) ensures that there exists a unique solution to the differential system and thus to the second order equation since all matrix elements have continuous real and imaginary parts in trading time $t\in[0,\tau_1]$.   Finally resubstitution leads to a unique solution to the Riccati-type equation.

\end{document}